\documentclass[reprint,onecolumn,showpacs,nofootinbib,prd,amssymb]{revtex4}
\usepackage{latexsym}
\usepackage{graphicx,epsf, epsfig, amssymb, amsmath}

\usepackage[usenames,dvipsnames]{xcolor}

\usepackage{relsize}
\usepackage{dsfont}
\usepackage{todonotes}
\usepackage{textcomp}
\usepackage{float}
\usepackage{color}

\newcommand{\aap}{Astron.\ Astrophys.}
\newcommand{\mnras}{Mon.\ Not.\ R.\ Astron.\ Soc.}
\newcommand{\physrep}{Phys.\ Rep.}
\newcommand{\apjl}{Astrophys.\ J.\ Lett.}

\newcommand{\apss}{Astrophys.\ Space Sci.}

\begin{document}

\title{Evolution of the magnetic field in neutron stars}

\author{M.~E.~Gusakov, E.~M.~Kantor, D.~D.~Ofengeim} 
\affiliation{Ioffe Physical-Technical Institute of the Russian Academy
  of Sciences, Polytekhnicheskaya 26, 194021 Saint-Petersburg, Russia}

\begin{abstract} 
We propose a general method 
to self-consistently study the quasistationary evolution
of the magnetic field in the cores of neutron stars.
The traditional approach to this problem is critically revised.
Our results are illustrated by calculation of the typical timescales
for the magnetic field dissipation as functions of temperature and the magnetic field strength.
\end{abstract}

\date{\today}



\maketitle

\section{Introduction}

Magnetic field plays a crucial role in the evolution of neutron stars (NSs).
Quite possibly, it 
serves as a most important unifying agent 
relating and explaining 
the observational properties of many diverse classes of NSs
(e.g., rotation-powered pulsars, magnetars, isolated neutron stars etc.) 
\cite{kaspi10,vigano_etal13}.
If this is true, non-accreting NSs from different classes 
differ mainly in their age 
and the magnetic field at birth.
To extract as much information from observations as possible
one, therefore, needs 
to be able to adequately model the long-term magneto-thermal evolution of NSs 
with different initial magnetic field configurations.
Clearly, this is a very complex theoretical problem, which has not been fully solved yet 
(but see, e.g., Refs.\ \cite{hui90, us95, pgz00, act04, apm08, pmg09, vigano_etal13}).

One of the aspects of this problem is the way magnetic field evolves
and dissipates in the internal layers of NSs.
Up to now a substantial body of research
on this subject has been concentrated on the crust 
(see, e.g., Refs.\ \cite{jones88, su97, rg02, hr04, gcr_etal13, gc14, gwh16} and references therein).
Because the ionic lattice of the crust
is immobile, the magnetic field there
evolves exclusively through the Ohmic decay and Hall drift.
The case of the core is much more complex, since there we have at least three particle species
(neutrons, protons, and electrons) that can move 
one relative to another,
so the diffusion effects come into play in addition to the two processes active in the crust.

Evolution of the magnetic field in the core has been studied, 
under various simplifying assumptions, e.g., in Refs.\ 
\cite{bpp69, hui90, pethick92, ur94, us95, td96, us99, kg00, kg01, 
act04, bs06, hrv08, dss09, hrv10, ho11, dgp12, gagl15, eprgv16, bl16}.
However, 
self-consistent
analysis of this problem has never been attempted.
To perform such an analysis one needs to solve (iterate in time) 
the Faraday induction equation, 
$\partial {\pmb B}/\partial t = -c \, {\pmb \nabla} \times {\pmb E}$,
where the electric field ${\pmb E}$ depends itself on the magnetic field ${\pmb B}$,
diffusion currents, 
perturbed chemical potentials etc.
A primary problem, therefore, 
consists in finding ${\pmb E}$ (and other parameters in the system) 
for a given quasistationary magnetic field configuration.
This problem has been addressed 
in a number of papers \cite{su95,td96,hrv08, reisenegger09, hrv10,gjs11,bl16,papm17} starting from the
work of Goldreich and Reisenegger \cite{gr92}.
Unfortunately, the validity of some approximations made in these references
remains unclear.
Here we reconsider this problem.
Namely, we propose a method 
of obtaining the
self-consistent solutions 
describing 
quasi-stationary evolution of the magnetic field in NSs.
Our results indicate that the conventional approach 
of Refs.\ \cite{gr92,td96,hrv08, reisenegger09, hrv10,gjs11,bl16,papm17}
may not be adequate.

The paper is organised as follows.
In Sec.\ \ref{maineq} 
we formulate dynamic equations
describing a magnetized mixture of nonsuperfluid/nonsuperconducting particles 
(e.g., neutrons, protons, and electrons in the NS core).
In Sec.\ \ref{perturb}
we propose a general scheme, allowing us to determine 
all the necessary ingredients to calculate the electric field in 
an NS with a specified (axisymmetric) magnetic field configuration.
In Sec.\ \ref{perturb5} we discuss how the proposed scheme should be modified
to account for muons (or other particle species), non-axisymmetric fields,
and nucleon superfluidity/superconductivity.
In Sec.\ \ref{dissip} we derive 
expressions for the 
dissipation rate of the magnetic energy in NS cores.
In Sec.\ \ref{example} the results of the preceding sections are illustrated by
calculation (and comparison) of the magnetic field decay rates due to different dissipation 
processes: Ohmic decay, non-equilibrium beta-reactions, and ambipolar diffusion.
Finally, Sec.\ \ref{concl} contains our conclusions and summary of results.

\section{General equations}
\label{maineq}

We consider a nonsuperfluid and nonsuperconducting matter 
composed of various (possibly, charged) particle species $\alpha$.
The effects of General relativity are neglected for clarity,%
%
\footnote{They do not affect our qualitative conclusions and can be easily incorporated.}
%
but the equation of state
is assumed to be fully relativistic.%
We also neglect thermal forces and 
the effects of temperature on the equation of state.
Then the equations that govern evolution of the system can be written as 
(see, e.g., Refs.\ \cite{ys91a,gr92,papm17})
\begin{align}
&\frac{\partial}{\partial t}\left( \frac{\mu_\alpha n_\alpha}{c^2} 
\, {\pmb u}_\alpha \right)
+\nabla_k \left( \frac{\mu_\alpha n_\alpha}{c^2} u_{(\alpha) i} u_{(\alpha) k} \right)&
\nonumber\\
&=\frac{\mu_{\alpha} \Delta \Gamma_\alpha}{c^2}  {\pmb u}_\alpha+
e_{\alpha} n_{\alpha} \left( {\pmb E}+\frac{1}{c}\left[{\pmb u}_{\alpha} \times {\pmb B} \right]\right)
-n_{\alpha} {\pmb \nabla}\mu_{\alpha} 
- \frac{\mu_{\alpha} n_{\alpha}}{c^2} \, {\pmb \nabla}\phi
-\sum_{\beta \neq \alpha} J_{\alpha\beta}({\pmb u}_{\alpha}-{\pmb u}_{\beta}),&
\label{main1}\\
&\frac{\partial n_\alpha}{\partial t}+{\rm div} (n_\alpha \pmb{u}_\alpha) = \Delta \Gamma_\alpha,&
\label{cont}\\
&\Delta \phi=\frac{4 \pi G}{c^2}\, (P+\varepsilon),&
\label{grav}\\
&\frac{\partial {\pmb B}}{\partial t} = -c \, {\pmb \nabla}\times {\pmb E},&
\label{max1}\\
&{\pmb \nabla}\times {\pmb B} = \frac{4 \pi}{c} \, {\pmb j},&
\label{rotB}\\
&\sum_{\alpha} e_\alpha n_\alpha=0.&
\label{quasi}
\end{align}
The first of these equations is the momentum conservation equation for each particle species $\alpha$;
the subscripts $i$ and $k$ there are the spatial indices.
The physical meaning of other equations is clear.
In these equations $c$ and $G$ are the speed of light and gravitational constant, respectively;
$\mu_{\alpha}$ and $n_{\alpha}$ 
are the relativistic chemical potential and number density for particle species $\alpha$;
$\phi$ is the gravitational potential;
$P$ and $\varepsilon$ are the total pressure and energy density, respectively,
\begin{align}
&P=-\varepsilon+\sum_{\alpha}\mu_{\alpha}n_{\alpha},&
\label{P}\\
&dP=\sum_{\alpha} n_{\alpha} d\mu_{\alpha};&
\label{dP2}
\end{align}
${\pmb u}_{\alpha} \equiv [u_{(\alpha)1},\, u_{(\alpha)2},\,u_{(\alpha)3}]$ 
is the velocity of particles $\alpha$;
$J_{\alpha\beta}=J_{\beta\alpha}$ is the quantity defined (and calculated) 
in Ref.\ \cite{ys91a} and related to the effective relaxation time $\tau_{\alpha\beta}$
for scattering of particles $\alpha$ on particles $\beta$ by the formula \cite{ys91a}:
$\tau_{\alpha\beta}=n_\alpha \mu_{\alpha}/(c^2 J_{\alpha\beta})$;
the terms in Eq.\ (\ref{main1}), which depend on  
$J_{\alpha\beta}$, 
represent the friction forces.
Further, ${\pmb E}$ and ${\pmb B}$ are the electric and magnetic fields;
$e_\alpha$ is the electric charge of particle species $\alpha$; and
\begin{align}
&{\pmb j}=\sum_{\alpha} e_\alpha n_{\alpha} {\pmb u}_{\alpha}&
\label{j}
\end{align}
is the electric current density 
[note that ${\rm div} \, {\pmb j}=0$ in view of Eq.\ (\ref{rotB})]. 
Finally, the source $\Delta \Gamma_\alpha$ in the continuity equation (\ref{cont})
appears due to non-equilibrium processes of particle mutual transformations 
(e.g., non-equilibrium Urca processes \cite{ykgh01}).
Note that we neglected the displacement current in Eq.\ (\ref{rotB}) and assumed 
the quasineutrality condition (\ref{quasi}), 
which is a perfect approximation for slow processes we are interested in
(see, e.g., Ref.\ \cite{ll60} for justification of this assumption).

The system of equations (\ref{main1})--(\ref{quasi}) is rather general but it is too complex.
It can be further simplified if the hydrodynamic description of the system 
is justified, i.e.\ if the inter-particle collisions are so frequent,
that $\tau_{\alpha\beta}\ll \tau_{\rm B}$ (see Sec.\ \ref{example}), 
where $\tau_{\rm B}$ is a typical timescale of the problem
(in our case, it is the magnetic field evolution timescale).
Then
the velocities ${\pmb u}_\alpha$ of different particle species $\alpha$ 
are very close to one another
(e.g., Ref.\ \cite{braginskii65}), so that it is 
convenient to
introduce the macroscopic velocity of the flow ${\pmb U}$ according to~\cite{ys91a}
\begin{align}
&{\pmb U} \sum_{\alpha} \mu_{\alpha} n_{\alpha} \equiv     
\sum_{\alpha} \mu_{\alpha}n_{\alpha} \, {\pmb u}_{\alpha},&
\label{222x}
\end{align}
and replace Eq.\ (\ref{main1}) with%
\begin{align}
&\frac{\partial}{\partial t}\left( \frac{\mu_\alpha n_\alpha}{c^2} 
\, {\pmb U} \right)
+\nabla_k \left( \frac{\mu_\alpha n_\alpha}{c^2} U_{i} U_{k} \right)&
\nonumber\\
&=\frac{\mu_{\alpha} \Delta \Gamma_\alpha}{c^2}  {\pmb U}
+e_{\alpha} n_{\alpha} \left( {\pmb E}+\frac{1}{c}\left[{\pmb u}_{\alpha} \times {\pmb B} \right]\right)
-n_{\alpha} {\pmb \nabla}\mu_{\alpha} 
- \frac{\mu_{\alpha} n_{\alpha}}{c^2} \, {\pmb \nabla}\phi
-\sum_{\beta \neq \alpha} 
J_{\alpha\beta}
\, ({\pmb u}_{\alpha}-{\pmb u}_{\beta}).&
\label{33x}
\end{align}
To obtain this equation we neglected a number of small terms in Eq.\ (\ref{main1}), 
making use of the fact that $\tau_{\rm B}\gg \tau_{\alpha\beta}=n_\alpha \mu_\alpha/(c^2 J_{\alpha\beta})$
(see, e.g., Ref.\ \cite{braginskii65} for a similar discussion).
For example, we neglected the terms 
$\partial/\partial t[\mu_\alpha n_{\alpha} ({\pmb u}_\alpha-{\pmb U})/c^2]\sim
\mu_\alpha n_{\alpha} ({\pmb u}_\alpha-{\pmb U})/(\tau_{\rm B} c^2)$
in comparison to the terms 
$\sum_{\beta\neq\alpha} J_{\alpha \beta} ({\pmb u}_{\alpha}-{\pmb u}_\beta)=
\sum_{\beta\neq\alpha} \mu_\alpha n_\alpha/(c^2 \tau_{\alpha\beta}) \, ({\pmb u}_{\alpha}-{\pmb u}_\beta)$.
We also replaced ${\pmb u}_\alpha$ with ${\pmb U}$ in (already) small dissipative term
$\mu_{\alpha}\Delta \Gamma_\alpha {\pmb u}_\alpha/c^2$, 
appearing due to the action of weak processes of particle mutual transformations.

Summing up Eq.\ (\ref{33x}) over all particle species 
and neglecting the term, quadratically small in the deviation from chemical equilibrium,
one obtains the standard force balance equation for the system as a whole,
\begin{equation}
\frac{\partial}{\partial t}\left[ \frac{(P+\varepsilon)}{c^2} 
\, {\pmb U} \right]
+\nabla_k \left[\frac{(P+\varepsilon)}{c^2} U_{i} U_{k} \right]
= \frac{1}{c}\left[{\pmb j} \times {\pmb B} \right]
-{\pmb \nabla}P
- \frac{(P+\varepsilon)}{c^2}\, {\pmb \nabla}\phi.
\label{44x}
\end{equation}
%

\section{The problem of magnetic field evolution in the NS cores: General scheme of the solution}
\label{perturb}

The equations of the previous section  describe an arbitrary mixture (plasma) of charged particles,
provided that they are nonsuperfluid and nonsuperconducting.
Here we apply them to the particular case of NS matter composed of 
neutrons ($n$), protons ($p$), and electrons ($e$) [$npe$-matter]. 
Extension of these results to more complex NS core compositions 
(e.g., $npe$-matter with an admixture of muons)
is straightforward and is discussed in Sec.\ \ref{perturb5},
where we also consider the effects of 
non-axisymmetric magnetic field and 
nucleon superfluidity/superconductivity.

\subsection{Our approximations and further simplifications}
\label{perturb2}

Assume that the star is nonrotating and spherically symmetric in the absence of the magnetic field.
It is in hydrostatic, diffusion, and beta-equilibrium;
all particle currents are absent.
Then we slightly perturb the system by creating some small currents. 
They generate the magnetic field, which we, for simplicity, take axisymmetric, 
${\pmb B}={\pmb B}(r, \, \theta)$ 
(non-axisymmetric case is briefly analysed in Sec.\ \ref{perturb5b}).%
%
\footnote{We do not consider here the question of stability of such system with respect to spontaneous 
reconfiguration of the magnetic field on the Alfven timescale.
The magnetic field configuration is assumed to be stable.
 }
%

After perturbation is applied, the system starts to evolve 
to equilibrium through particle diffusion and beta-processes. 
This process is accompanied by the magnetic field dissipation.
A typical timescale for reaching the equilibrium (i.e., the timescale of magnetic field decay)
is very large (see below), 
so that the system evolves through a set of quasistationary states,
which means that one can neglect time derivatives in Eqs.\ (\ref{main1}), (\ref{cont}),
(\ref{33x}), (\ref{44x}) of the previous section
[and, in addition, ignore the quadratically small velocity-dependent terms, in particular,
the term depending on $U_i U_k$ 
in Eq.\ (\ref{44x})].
We follow here the ideas of Ref.\ \cite{gr92}.

Let us demonstrate, for example, that the time derivative in 
the continuity equation (\ref{cont}) can be omitted.
Below the perturbation of a quantity $A$ will be denoted as $\delta A$.
From Eqs.\ (\ref{rotB}) and (\ref{44x}) it follows that
the perturbation of the pressure $P$ by the magnetic field is
$\delta P \sim B^2$. 
Correspondingly, $\delta n_{\alpha} \sim n_{\alpha} B^2/P$
and $\partial n_{\alpha}/\partial t \sim n_{\alpha} B^2/(P \tau_{\rm B})$.
The magnetic evolution timescale $\tau_{\rm B}$
is given by [see Eq.\ (\ref{max1})]: $B/\tau_{\rm B} \sim c E/R \sim u_{\alpha}B/R$,
hence $\tau_{\rm B}\sim R/u_{\alpha}$
(here $R$ is the typical lengthscale; 
to obtain $\tau_{\rm B}$ we estimated $E$ as: $E\sim u_{\alpha} B/c$).
Now we can write $\partial n_{\alpha}/\partial t \sim n_{\alpha} B^2/(P \tau_{\rm B}) \sim
n_\alpha u_\alpha B^2/(P R)$,
while $|{\rm div}(n_{\alpha} {\pmb u}_{\alpha})| \sim n_{\alpha} u_{\alpha}/R$.
Comparing these terms, 
it is easy to see that
$\partial n_{\alpha}/\partial t$
drops out from the continuity equation written to leading order in $B^2/P$.

Accounting for the 
approximations listed above, 
Eq.\ (\ref{44x}) can be represented as 
[we make use of Eqs.\ (\ref{P}) and (\ref{dP2})]
\begin{equation}
\sum_{\alpha} n_{\alpha} {\pmb \nabla}\mu_{\alpha}^{\infty}
  =\frac{1}{c}\left[{\pmb j} \times {\pmb B} \right],
\label{44xx}
\end{equation}
where we introduced the redshifted chemical potentials,
$\mu_{\alpha}^{\infty} = \mu_{\alpha} e^{\phi/c^2}$;
in the weak-field approximation
${\pmb \nabla}\mu_{\alpha}^{\infty} \approx {\pmb \nabla}\mu_{\alpha}+
(\mu_{\alpha}/c^2) \, {\pmb \nabla}\phi$.
Using the quasineutrality condition (\ref{quasi}) and the definitions 
$\Delta \mu_e \equiv \mu_p+\mu_e-\mu_n$ and $n_b=n_p+n_n$,
Eq.\ (\ref{44xx}) can be rewritten as
\begin{equation}
n_{e} {\pmb \nabla}\Delta \mu_e^{\infty} + n_b {\pmb \nabla} \mu^{\infty}_n
=\frac{1}{c}\left[{\pmb j} \times {\pmb B} \right],
\label{444xx}
\end{equation}
In full thermodynamic and hydrostatic equilibrium (when there is no magnetic field) 
one has 
\begin{align}
\Delta \mu_e^{\infty}=\Delta \mu_e=0, \quad \quad \mu_n^{\infty}={\rm const}.
\label{hydr}
\end{align}
When the magnetic field is applied, there is a small deviation from equilibrium, and
\begin{equation}
n_{e} {\pmb \nabla}\delta \Delta \mu_e^{\infty} + n_b {\pmb \nabla} \delta \mu^{\infty}_n
=\frac{1}{c}\left[{\pmb j} \times {\pmb B} \right].
\label{4444xx}
\end{equation}
Taking into account that, in view of Eq.\ (\ref{hydr}),
\begin{align}
\delta  \Delta\mu_e^\infty =\Delta \mu_e^\infty \approx \Delta \mu_e \left(1+\frac{\phi}{c^2} \right)
\approx \Delta \mu_e,
\label{dm1}
\end{align}
one obtains that, 
to leading order in the deviation, 
Eq.\ (\ref{4444xx}) can be represented as
\begin{equation}
{\pmb \nabla } \left( 
n_{e} \Delta \mu_e + n_b \delta \mu^{\infty}_n
\right)- \left(
\frac{dn_e}{dr} \Delta \mu_e +\frac{dn_b}{dr} \delta \mu^{\infty}_n \right) {\pmb e}_r
=\frac{1}{c}\left[{\pmb j} \times {\pmb B} \right],
\label{44444xx}
\end{equation}
where the functions $n_e(r)$ and $n_b(r)$ can be thought of as taken in equilibrium
and ${\pmb e}_r$ is the unit vector in radial direction.
The left-hand side of Eq.\ (\ref{44444xx}) depends on two scalars 
determined by the functions 
$\Delta \mu_e(r,\, \theta)$ and $\delta \mu_n^{\infty}(r,\, \theta)$.
It turns out that 
${\pmb B}(r,\, \theta)$ in this situation 
cannot be arbitrary in order to compensate 
the left-hand side of Eq.\ (\ref{44444xx}).
At the very least, for axisymmetric fields, 
the $\varphi$-component of the Lorentz force density, 
${\pmb F}_{\rm L}=[{\pmb j}\times {\pmb B}]/c$, 
must vanish 
(gradient of an axisymmetric function cannot have non-zero $\varphi$-component),
\begin{align}
F_{\rm L \varphi}=\frac{1}{c}\, [{\pmb j}\times {\pmb B}]_{\varphi}=0.
\label{Lorentz}
\end{align}
As shown, e.g., in Refs.\ \cite{reisenegger07,gl16}, 
this is 
the only 
constraint imposed on the magnetic field in order to satisfy (\ref{44444xx}).
Then both functions $\Delta \mu_e(r,\, \theta)$ 
and $\delta \mu_n^{\infty}(r,\, \theta)$ can be expressed through ${\pmb F}_{\rm L}$
and some unknown 
scalar function $\zeta(r)$ (see Appendix \ref{A1}),
which will be determined
in Sec.\ \ref{perturb44a}.
Below in this section and in Sec.\ \ref{perturb1} 
we assume that $\Delta \mu_e(r,\, \theta)$ and  $\delta \mu_n^{\infty}(r,\, \theta)$ are already 
found.
Then, in the weak-field limit, 
$\delta \mu_n$  
is given by
\begin{align}
\delta  \mu_n \approx  \delta \mu_n^\infty-\frac{\mu_{n0}}{c^2} \, \delta \phi,
\label{dm2}
\end{align}
where 
$\mu_{n0}$ 
is the equilibrium function $\mu_{n}$. 
If we work in the Cowling approximation (i.e., assume $\delta \phi=0$),
this
function can be obtained immediately; 
otherwise, one should first determine 
the gravitational potential perturbation $\delta \phi$ from the Poisson's equation (\ref{grav}).
All in all, $\Delta \mu_{e}$ and $\delta\mu_{n}$ (and thus $\mu_{n}$)
can be determined.
This means that we know {\it any} thermodynamic quantity in the perturbed $npe$-matter, 
since it can be presented as a function of only 
three parameters,
e.g., $\Delta \mu_e$, $\mu_n$, and $T$
(we remind the reader that the quasineutrality condition, $n_e=n_p$, 
still holds true in the perturbed matter). 

At this stage our consideration starts to differ from 
that of Ref.\ \cite{gr92} and others (e.g., \cite{td96,hrv08,reisenegger09,hrv10,gjs11,bl16,papm17}).
In those references $\Delta \mu_e$ is determined from the 
{\it scalar} differential equation 
[see, e.g., equation (14) in Ref.\ \cite{papm17}], 
which is a divergence of a combination of the momentum equations (\ref{33x})
and the continuity equations (\ref{cont}).
This scalar equation is derived under simplifying assumptions,
whose validity for stratified  matter is questionable.
Moreover, the solution to this equation 
is not necessary a solution to the initial vector Eqs.\ (\ref{cont}) and (\ref{33x}).
As a result, $\Delta \mu_e$ 
in Refs.\ \cite{gr92, td96,hrv08,reisenegger09,hrv10,gjs11,bl16,papm17} depends on 
the rate 
of beta-processes and on the relaxation time $\tau_{np}$,
which
are sensitive functions of temperature $T$ (e.g., Refs.\ \cite{ykgh01, ys91a}).
In contrast, here we argue that $\Delta \mu_e$ and $\mu_n$ 
do not depend on $T$ and
are fixed 
by the magnetic field configuration.%
%
\footnote{More precisely, if we expand $\Delta \mu_e$ or $\mu_n$ 
in a series of Legendre polynomials $P_l({\rm cos} \theta)$,
then all the components except for $l=0$ will be independent of temperature;
the $l=0$ component 
may vary with temperature, 
but only in a narrow temperature range 
(see Sec.\ \ref{perturb44a} and Appendix \ref{A3} for more details).}
%
Further critical analysis of the previous results on the subject 
is presented in Appendix \ref{critique}.

\subsection{Determining the velocities ${\pmb u}_\alpha$ in the comoving frame}
\label{perturb1}

To study the magnetic field evolution in NS cores it is
necessary to extract all the available information from the dynamic equations discussed above.
Here our aim will be to find the velocities ${\pmb u}_\alpha$.
We shall work in the \underline{locally comoving} coordinate system, in which ${\pmb U}=0$.
In that coordinate system we define the vectors ${\pmb \nu}_{\alpha}$,
\begin{align}
{\pmb u}_{\alpha\, {\rm comoving}} \equiv {\pmb \nu}_{\alpha}/n_\alpha,
\label{coord}
\end{align}
where ${\pmb u}_{\alpha\, {\rm comoving}}$ is the velocity of particle species $\alpha$
in the comoving frame.
Correspondingly, in an arbitrary frame
\begin{align}
n_\alpha {\pmb u}_{\alpha} \equiv n_{\alpha} {\pmb U}+ {\pmb \nu}_{\alpha}.
\label{coord2}
\end{align}
To find ${\pmb \nu}_\alpha=n_\alpha {\pmb u}_{\alpha \, {\rm comoving}}$ 
we should use Eq.\ (\ref{33x}) which reduces, in our problem, to %
%
\footnote{Note that, only two of the three Eqs.\ (\ref{33x}) for neutrons, protons, and electrons 
are really independent, since they contain Eq.\ (\ref{444xx}) that has already been used.}
%
%
\begin{equation}
0=
e_{\alpha} n_{\alpha} \left( {\pmb E}_{\rm comoving}
+\frac{1}{c}\left[{\pmb u}_{\alpha\, {\rm comoving}} \times {\pmb B} \right]\right)
-n_{\alpha} {\pmb \nabla}\mu_{\alpha}^{\infty} 
-\sum_{\beta \neq \alpha} 
J_{\alpha\beta}
\, ({\pmb u}_{\alpha \, {\rm comoving}}-{\pmb u}_{\beta \, {\rm comoving}}),
\label{333x}
\end{equation}
together with the condition ${\pmb U}=0$, which is equivalent to 
[see the definition (\ref{222x})]
\begin{equation}
\sum_{\alpha} \mu_{\alpha}n_{\alpha} \, {\pmb u}_{\alpha \, {\rm comoving}}
\equiv \sum_{\alpha}\mu_\alpha {\pmb \nu}_\alpha=0.
\label{2222x}
\end{equation}
In Eq.\ (\ref{333x}) ${\pmb E}_{\rm comoving}$ is the electric field 
in the comoving frame.
It is related to the electric field ${\pmb E}$ in the laboratory frame
by the formula
\begin{align}
{\pmb E}_{\rm comoving}={\pmb E}+\frac{1}{c}\, [{\pmb U}\times {\pmb B}].
\label{EcomE}
\end{align}

A solution to the system of linear equations (\ref{333x}) and (\ref{2222x})
will give us ${\pmb \nu}_{n}$, ${\pmb \nu}_{p}$, and ${\pmb \nu}_e$ as functions
of ${\pmb \nabla}\mu_\alpha^\infty$ ($\alpha=n$, $p$, $e$), ${\pmb E}_{\rm comoving}$, and ${\pmb B}$.
[We remind the reader that we assume (see Sec.\ \ref{perturb2}) 
that all the perturbed thermodynamic quantities, in particular, 
${\pmb \nabla}\mu_\alpha^\infty$, are already ``calculated''.]
The resulting expressions are rather lengthy so here we present them schematically,
\begin{align}
{\pmb \nu}_\alpha={\pmb \nu}_\alpha({\pmb \nabla}\mu_\beta^\infty, \, {\pmb E}_{\rm comoving},\, {\pmb B}).
\label{nu222}
\end{align}
The (unknown) electric field ${\pmb E}_{\rm comoving}$ 
can then be determined from the condition [see Eq.\ (\ref{j})]
\begin{align}
{\pmb j}=e_e {\pmb \nu}_e+e_p {\pmb \nu}_p,
\label{jcond}
\end{align}
so that (again schematically) ${\pmb E}_{\rm comoving}$ is given by
\begin{align}
{\pmb E}_{\rm comoving}={\pmb E}_{\rm comoving}({\pmb \nabla}\mu_\alpha^\infty, \, {\pmb B}, \, {\pmb j})
\label{Ecom}
\end{align}
and hence [see Eq.\ (\ref{EcomE})]
\begin{align}
{\pmb E}={\pmb E}_{\rm comoving}({\pmb \nabla}\mu_\alpha^\infty, \, {\pmb B}, \, {\pmb j})
-\frac{1}{c} [{\pmb U}\times {\pmb B}].
\label{Ecom2}
\end{align}
Now, substituting Eq.\ (\ref{Ecom}) into (\ref{nu222}), 
the currents ${\pmb \nu}_\alpha$ can be represented as only functions of 
${\pmb \nabla}\mu_\beta^\infty$ ($\beta=n$, $p$, $e$), ${\pmb B}$, 
and ${\pmb j}=\frac{c}{4\pi}\, {\pmb \nabla}\times{\pmb B}$,
\begin{align}
{\pmb \nu}_\alpha = {\pmb \nu}_\alpha({\pmb \nabla}\mu_\beta^\infty, \, {\pmb B}, \, {\pmb j}).
\label{nui}
\end{align}
In fact, 
these functions can be found analytically.
We also emphasize that the quantities ${\pmb \nabla}\mu_\beta^\infty$ and ${\pmb j}$
are itself determined by the magnetic field; 
the gradients ${\pmb \nabla}\mu_\beta^\infty$, in addition, depend
on the unknown function $\zeta(r)$ 
(see Sec.\ \ref{perturb2} and Appendix \ref{A1}),
which will be determined in the next section.

\subsection{Determination of the flow velocity ${\pmb U}$ and the function $\zeta(r)$}
\label{perturb44}

Our next (most important) 
step will be to determine the flow velocity ${\pmb U}$ 
and the function $\zeta(r)$,
the only unknown parameters remained.

\subsubsection{The components $U_r$ and $U_\theta$ and the equation for $\zeta(r)$}
\label{perturb44a}

To this aim, let us consider the continuity equations (\ref{cont}).
They can now be rewritten as [see the definition (\ref{coord2})]
\begin{align}
&{\rm div} (n_{e} {\pmb U})=-{\rm div}\, {\pmb \nu}_{e}-\Delta \Gamma,&
\label{cont2e}\\
&{\rm div} (n_{p} {\pmb U})=-{\rm div} \,{\pmb \nu}_{p}-\Delta \Gamma,&
\label{cont2p}\\
&{\rm div} (n_{n} {\pmb U})=-{\rm div} \, {\pmb \nu}_{n}+\Delta \Gamma,&
\label{cont2n}
\end{align}
where $\Delta\Gamma \equiv \Delta\Gamma_n=-\Delta\Gamma_p=-\Delta\Gamma_e$
(note that $\Delta\Gamma$ is the well-known function of $\Delta \mu_e$ and $T$ \cite{ykgh01}).
Because $n_e=n_p$, 
Eq.\ (\ref{cont2e}) is duplicate of (\ref{cont2p}) and can be omitted.%
%
\footnote{It may seem that these equations contain one more non-trivial condition, 
${\rm div}\, {\pmb \nu}_e={\rm div} \, {\pmb \nu}_p$. 
But this condition means ${\rm div} \, {\pmb j}=0$,
which is satisfied ``by construction''  (automatically) in view of Eq.\ (\ref{rotB}).}
%
The remaining Eqs.\ (\ref{cont2p}) and (\ref{cont2n}) can be rewritten,
to leading order in the deviation from equilibrium, as
\begin{align}
&n_p \, {\rm div} \,{\pmb U} + \frac{d n_p}{dr} \, U_r =-{\rm div} \,{\pmb \nu}_{p}-\Delta \Gamma,&
\label{cont2p2}\\
&n_n \, {\rm div} \, {\pmb U} + \frac{d n_n}{dr} \, U_r =-{\rm div} \,{\pmb \nu}_{n}+\Delta \Gamma.&
\label{cont2n2}
\end{align}
These equations can be solved for ${\rm div} \, {\pmb U}(r,\, \theta)$ and $U_r(r, \, \theta)$,%
%
\footnote{
\label{nonstrat}
Note that, the solution does not exist for a non-stratified star.
Then it is possible to modify our scheme in order to determine ${\pmb U}$ 
(and other quantities of interest).
However, we prefer not to discuss this unrealistic case in the paper.}
%
then $U_\theta(r,\, \theta)$ can be easily found, 
\begin{align}
U_\theta=\frac{1}{{\rm sin} \theta}\left\{ 
\int_0^\theta r \, {\rm sin} \tilde{\theta} \left[ {\rm div} \, {\pmb U} - \frac{1}{r^2} 
\frac{\partial(r^2 U_r)}{\partial r}
\right] d \tilde{\theta} + \xi(r)
\right\},
\label{Uth}
\end{align}
where $\xi(r)$ is some function which must vanish, $\xi(r)=0$, 
to guarantee finiteness of $U_\theta$ at $\theta=0$.
Another potentially dangerous point where $U_\theta$ 
can be infinite corresponds to $\theta=\pi$.
The condition ensuring that it is not the case reads
\begin{align}
\int_0^\pi r \, {\rm sin} \tilde{\theta} \left[ {\rm div} \, {\pmb U} - \frac{1}{r^2} 
\frac{\partial(r^2 U_r)}{\partial r}
\right] d \tilde{\theta}=0.
\label{Uthetacond}
\end{align}
This equation indicates
that the multipole $l=0$
in the Legendre expansion of the function in the square brackets
must vanish.
That function depends on the chemical potentials $\Delta \mu_e$ and $\mu_n^\infty$
and hence on (still unknown) function $\zeta(r)$ introduced in Appendix \ref{A1}.
The condition (\ref{Uthetacond}), therefore, 
should be considered as a {\it differential} equation for $\zeta(r)$;
it should be supplied by the
boundary conditions, which 
follow, in particular, 
from the requirement of the regularity of $U_{r}$ at $r \rightarrow 0$,
and are discussed in more detail in Appendix \ref{A3}.
A solution of Eq.\ (\ref{Uthetacond}) allows us to find $\zeta(r)$ and hence to fully determine 
the quantities $\Delta \mu_e$ and $\mu_n^\infty$, 
as has already been advertised in Sec.\ \ref{perturb2}.

\subsubsection{The component $U_\varphi$}
\label{perturb44b}

And what about $U_\varphi$? 
It does not enter the dynamic equations described above, 
except for the magnetic field evolution equation (\ref{max1}), 
which can be rewritten as [see Eq.\ (\ref{Ecom2})]
\begin{align}
&\frac{\partial {\pmb B}}{\partial t} =-c {\pmb \nabla}\times 
{\pmb E}=
-c \, {\pmb \nabla}\times 
{\pmb E}_{\rm comoving}({\pmb \nabla}\mu_\alpha^\infty, \, {\pmb B}, \, {\pmb j})
+{\pmb \nabla}\times [{\pmb U}\times {\pmb B}].&
\label{max2}
\end{align}
How could we determine it?
The idea is to look more carefully at the force balance equation (\ref{4444xx}).
Assume that, initially, our system is quasistationary, that is
Eqs.\ (\ref{44444xx}) and (\ref{Lorentz}) are satisfied.
After a short (in comparison to the diffusion timescale) period of time $\delta t$
the magnetic field will change according to Eq.\ (\ref{max2}),
\begin{align}
\delta {\pmb B}=\left\{-c \, {\pmb \nabla}\times 
{\pmb E}_{\rm comoving}({\pmb \nabla}\mu_\alpha^\infty, \, {\pmb B}, \, {\pmb j})
+{\pmb \nabla}\times [{\pmb U}\times {\pmb B}]\right\}\, \delta t.
\label{max3}
\end{align}
This will, in turn, change the Lorentz force density
by 
$\delta {\pmb F}_{\rm L} = [\delta {\pmb j} \times {\pmb B}]/c+[{\pmb j} \times \delta{\pmb B}]/c$.
The $r$- and $\theta$-components of $\delta {\pmb F}_{\rm L}$ can be easily 
compensated by adjusting the chemical potentials.
However, \underline{there is no compensating force along the $\varphi$-component}.
This means that $U_{\varphi}$
will be rapidly generated and become of the order of ${\pmb u}_{\alpha}$ 
on the Alfven timescale $t_{\rm A}\sim [\mu_n n_b R^2/(B^2 c^2)]^{1/2}
\sim \, 0.2 R_6^2/B_{14}^2$~s 
[this estimate follows from Eq.\ (\ref{44x})]. 
Eventually,
the system will evolve in a quasistationary manner with 
$\delta F_{\rm L\varphi}=0$ at each time step.
Mathematically, this amounts to an additional constraint,
\begin{align}
\frac{\partial F_{\rm L\varphi}}{\partial t}
=\frac{1}{c} 
\left(
\frac{\partial {\pmb j}}{\partial t}\times {\pmb B}
+{\pmb j}\times\frac{\partial {\pmb B}}{\partial t}
\right)_\varphi
= -\frac{c}{4\pi} 
\left( {\rm rot}\,{\rm rot}{\pmb E}\times{\pmb B}
+ {\rm rot}{\pmb B}\times {\rm rot}{\pmb E} \right)_\varphi=0,
\label{uphi}
\end{align}
where the electric field ${\pmb E}$ is given (schematically) by Eq.\ (\ref{Ecom2}).
This condition determines $U_{\varphi}$ and is necessary for quasistationarity
of the system.

\section{Various extensions: Accounting for muons, non-axisymmetric magnetic field, superfluidity/superconductivity, and deviations from the diffusion and beta-equilibrium, which are not related to the magnetic field }
\label{perturb5}

\subsection{Muons}
\label{perturb5a}

The scheme described above can be easily generalized to the case of $npe\mu$ matter
(an inclusion of other particle species, e.g., hyperons, is similar).
The force balance equation (\ref{44444xx}) in $npe\mu$-matter 
takes the form
\begin{equation}
{\pmb \nabla } \left( 
n_{e} \Delta \mu_e + n_{\mu} \Delta \mu_\mu+ n_b \delta \mu^{\infty}_n
\right)- \left(
\frac{dn_e}{dr} \Delta \mu_e +\frac{dn_\mu}{dr} \Delta \mu_\mu 
+\frac{dn_b}{dr} \delta \mu^{\infty}_n \right) {\pmb e}_r
=\frac{1}{c}\left[{\pmb j} \times {\pmb B} \right],
\label{fb}
\end{equation}
where $\Delta \mu_\mu \equiv \mu_p+\mu_\mu-\mu_n$; 
$\mu_\mu$ and $n_\mu$ are the muon chemical potential and number density, respectively.
Solution to this equation allows one to express, e.g., 
$\Delta \mu_e$ and $\delta \mu^{\infty}_n$ through the magnetic field
${\pmb B}$, the imbalance $\Delta \mu_\mu$, and the unknown function $\zeta(r)$.
Additional equation, which is necessary to determine $\Delta \mu_\mu$
is provided by the continuity equation for muons, 
\begin{align}
&{\rm div} (n_{\mu} {\pmb U})=-{\rm div}\, {\pmb \nu}_{\mu}-\Delta \breve{\Gamma},&
\label{contmu}
\end{align}
where $\Delta \breve{\Gamma}$ is the 
source (depending on $\Delta \mu_\mu$, $T$ and $n_b$) 
appearing due to non-equilibrium beta-processes involving muons
and the vector ${\pmb \nu}_\mu$ (which depends on $\Delta \mu_\mu$) 
has the same meaning as the vectors ${\pmb \nu}_\alpha$
from the preceding section; it can be found from the momentum equation for muons 
[analogous to Eq.\ (\ref{33x})].
The function $\zeta(r)$ should be determined from the requirement of regularity
of the solution for ${\pmb U}$ in the same way as it is done in Sec.\ \ref{perturb44a}.

\subsection{Non-axisymmetric magnetic field}
\label{perturb5b}

The case of non-axisymmetric magnetic field 
${\pmb B}={\pmb B}(r,\, \theta,\, \varphi)$ 
is of course much more 
complex, but the general scheme of Secs.\ \ref{perturb2}--\ref{perturb44}
remains applicable to that case as well.
The main difference concerns 
the constraint (\ref{Lorentz}) on the admissible configurations 
of the magnetic field. 
It is straightforward to show \cite{gl16} that in the non-axisymmetric case
it should be modified,
\begin{align}
\frac{\partial F_{\rm L \theta}}{\partial \varphi} =\frac{\partial}{\partial \theta}
\left({\rm sin} \theta \, F_{\rm L \varphi}\right).
\label{magncond}
\end{align}
Most of other equations 
[in particular, Eqs.\ (\ref{cont2p}) and (\ref{cont2n})] 
remain unchanged,
but the solution 
(\ref{Uth})
and the constraint (\ref{uphi})
should be disregarded.
Using Eq.\ (\ref{magncond}) 
and following the same line of reasoning as in Sec.\ \ref{perturb44},
it is easy to verify that, in the non-axisymmetric case, 
the constraint (\ref{uphi}) should be replaced with
\begin{align}
\frac{\partial}{\partial t}
\left[\frac{\partial F_{\rm L \theta}}{\partial \varphi} -\frac{\partial}{\partial \theta}
\left({\rm sin} \theta \, F_{\rm L \varphi}\right)\right]=0.
\label{constr}
\end{align}
Together with the continuity equations (\ref{cont2p}) and (\ref{cont2n}),
this constraint will allow one to determine the velocity ${\pmb U}$.
Note that Eq.\ (\ref{constr}) reduces to (\ref{uphi}) in the axisymmetric case.

\subsection{Superfluidity/superconductivity}
\label{perturb5c}

The general scheme considered in the above sections can also be applied
to superfluid and superconducting matter.
Consider, for example, $npe\mu$-matter in a {\it non-rotating} magnetized star, 
in which neutrons are superfluid at $T<T_{{\rm c}n}$ 
($T_{{\rm c}n}$ is the neutron critical temperature)
and protons are normal.
This situation has recently been considered in Ref.\ \cite{kg17} and we refer 
the interested reader to that reference for details.

In the presence of superfluidity the total force balance equation (\ref{fb})
retains its form, however, it should be supplemented by an additional constraint,
following from the superfluid equation for neutrons \cite{kg17},
\begin{align}
{\pmb\nabla} \mu_n^{\infty}= 
{\pmb \nabla} 
\delta \mu_n^{\infty}=0.
\label{mun}
\end{align}
Using it, one can easily 
express
(similarly to how it is done in Appendix \ref{A1}) 
the imbalances 
$\Delta \mu_e$ and $\Delta \mu_\mu$ from Eq.~(\ref{fb})
through 
the magnetic field and the function $\zeta(r)$ (to be determined below).%
%
\footnote{We remind that in beta-equilibrium $\Delta \mu_e=\Delta \mu_\mu=0$.}
%
Since we ``know'' $\mu_n$, 
$\Delta \mu_e$, and $\Delta \mu_\mu$, 
we can calculate any thermodynamic quantity in $npe\mu$-matter.

The next step is to employ the quasistationary Euler-type equations 
for electrons, muons, and protons.
They have a standard form (see Sec.\ \ref{maineq}),
\begin{eqnarray}
-e(\pmb E + \frac{1}{c} \pmb u_e \times \pmb B) - \pmb \nabla \mu_e^\infty 
-\frac{J_{ep}}{n_e}(\pmb u_e - \pmb u_p) -\frac{J_{en}}{n_e} (\pmb u_e - \pmb u_n)-\frac{J_{e\mu}}{n_e} (\pmb u_e - \pmb u_\mu)=0, 
\label{esflmu}\\
-e(\pmb E + \frac{1}{c} \pmb u_\mu \times \pmb B) - \pmb \nabla \mu_\mu^\infty 
-\frac{J_{\mu p}}{n_\mu}(\pmb u_\mu - \pmb u_p) -\frac{J_{\mu n}}{n_\mu} (\pmb u_\mu - \pmb u_n)-\frac{J_{e\mu}}{n_\mu} (\pmb u_\mu - \pmb u_e)=0, 
\label{musflmu}\\
e (\pmb E + \frac{1}{c} \pmb u_p \times \pmb B) -\pmb \nabla  \mu_p^\infty 
-\frac{J_{ep}}{n_p}(\pmb u_p - \pmb u_e) -\frac{J_{np}}{n_p} (\pmb u_p - \pmb u_n)-\frac{J_{\mu p}}{n_p} (\pmb u_p - \pmb u_\mu)=0,
\label{psflmu}
\end{eqnarray}
where ${\pmb u}_\mu$ is the muon velocity and 
${\pmb u}_{n}$ is the velocity of neutron thermal excitations.
Generally, it differs from the neutron superfluid ``velocity'', 
proportional to the gradient
of the phase $\Phi_n$ of the Cooper-pair condensate wave function (see below).
${\pmb u}_{n}$ can be expressed through the velocities 
${\pmb u}_e$, ${\pmb u}_\mu$, ${\pmb u}_p$ from the equation 
\begin{align}
J_{\mu n} (\pmb u_\mu - \pmb u_n)+J_{en} (\pmb u_e - \pmb u_n)+J_{np} (\pmb u_p - \pmb u_n)=0,
\label{velrelmu}
\end{align}
which follows \cite{kg17} 
from a combination of Eqs.\ (\ref{fb}) and (\ref{mun})--(\ref{psflmu}).%
%
\footnote{
\label{f1}
Note that only five of six Eqs.\ (\ref{fb}), (\ref{mun}), and (\ref{esflmu})--(\ref{velrelmu})
are really independent.
}
%
These equations should be supplemented by the definition 
of the charge current density, 
\begin{align}
{\pmb j}= \frac{c}{4\pi} \, {\pmb \nabla}\times {\pmb B} 
=\sum_{\alpha=\mu,\, e,\, p} e_\alpha n_\alpha {\pmb u}_{\alpha}.
\label{j2}
\end{align}

To proceed further, 
we define the macroscopic velocity ${\pmb U}$ of the flow of the normal component
(i.e., electrons, muons, protons, and neutron thermal excitations)
according to the condition
\begin{align}
{\pmb U} \left[\sum_{\alpha=e,\, p,\, \mu} \mu_{\alpha} n_{\alpha} + \mu_n n_{n, {\rm th}}
\right]
\equiv     
\sum_{\alpha= e,\, p,\, \mu} \mu_{\alpha}n_{\alpha} \, {\pmb u}_{\alpha}
+\mu_n n_{n,{\rm th}} \, {\pmb u}_n,
\label{Usfl}
\end{align}
where $n_{n,{\rm th}} \equiv n_n-\mu_n Y_{nn}$ is the 
number density of (normal) neutron thermal excitations
and $Y_{nn}$ is the  $nn$ component of the relativistic entrainment matrix 
\cite{ga06, gkh09a, gkh09b, gusakov16,gd16} 
(all other components of this matrix vanish when protons are normal).
It vanishes at $T>T_{{\rm c} n}$, $Y_{nn}=0$, 
and equals $Y_{nn}=n_n/\mu_n$ at $T=0$.
In the non-relativistic limit $Y_{nn}$ is related 
to the neutron superfluid density, $\rho_{{\rm s}n}$,
by $Y_{nn}=\rho_{{\rm s}n}/(m_n^2 c^2)$.

Now, working in the locally comoving frame (${\pmb U}=0$) 
and using Eqs.\ (\ref{musflmu})--(\ref{Usfl}) [Eq.\ (\ref{esflmu}) is ignored since
it is a linear combination of
other equations, see footnote \ref{f1}],
one can express the quantities 
${\pmb E}_{\rm comoving}$, ${\pmb \nu}_{e}$, ${\pmb \nu}_{\mu}$, ${\pmb \nu}_{p}$, and 
${\pmb \nu}_{n} \equiv n_{n,{\rm th}} ({\pmb u}_{n}-{\pmb U})$
through ${\pmb \nabla} \mu_\beta^\infty$, ${\pmb B}$, and ${\pmb j}$
in exactly the same way as it is done in Sec.\ \ref{perturb1} 
(the notation is the same as in that section).
The electric field ${\pmb E}$ in the laboratory frame is then given by Eq.\ (\ref{Ecom2}) and
depends on ${\pmb U}$.
To find ${\pmb U}$, one should employ the continuity equations,
\begin{align}
&{\rm div} (n_{e} {\pmb U})=-{\rm div}\, {\pmb \nu}_{e}-\Delta \Gamma,&
\label{cont3e}\\
&{\rm div} (n_{\mu} {\pmb U})=-{\rm div}\, {\pmb \nu}_{\mu}-\Delta \breve{\Gamma},&
\label{cont3mu}\\
&{\rm div} (n_{p} {\pmb U})=-{\rm div} \,{\pmb \nu}_{p}-\Delta \Gamma-\Delta \breve{\Gamma},&
\label{cont3p}\\
&{\rm div} 
\left[n_{n,{\rm th}} \, {\pmb U}+Y_{nn} c^2\, {\pmb \nabla} \left(\frac{\hbar \Phi_n}{2} \right)
\right]=-{\rm div} \, {\pmb \nu}_{n}+\Delta \Gamma+\Delta \breve{\Gamma},&
\label{cont3n}
\end{align}
where 
the second term in the left-hand side of Eq.\ (\ref{cont3n}) 
describes the motion of the superfluid neutron component 
(see, e.g., Refs.\ \cite{ga06, gusakov16, gd16}).

As in Sec.\ \ref{perturb44},
one of the equations (\ref{cont3e})--(\ref{cont3p}) [e.g., Eq.\ (\ref{cont3e})] 
can be disregarded because of the quasineutrality condition, $n_p=n_e+n_\mu$,
and charge conservation, ${\rm div} \, {\pmb j}=0$.
Then the components $U_r$ and $U_\theta$ of the velocity ${\pmb U}$ 
can be found from Eqs.\ (\ref{cont3mu}) and (\ref{cont3p});
the function $\zeta(r)$ follows from the differential equation
ensuring regularity of $U_r$ and $U_\theta$.
The component $U_{\varphi}$ is still given by the condition (\ref{uphi}), 
which retains its form in the superfluid $npe\mu$-matter
provided that the magnetic field is axisymmetric.
Finally, the neutron continuity equation allows one to determine the phase $\Phi_n$
of the wave function of the Cooper-pair condensate.
Thus, all the unknown parameters in the system can be found
following the same strategy as in Sec.\ \ref{perturb}.

In principle, these results can be extended to account for proton superconductivity.
In particular, the total force balance equation will take the form 
[for $npe$-matter, cf. Eq.\ (\ref{44444xx})]
\begin{align}
{\pmb \nabla } \left( 
n_{e}  \Delta \mu_e + n_b \delta \mu^{\infty}_n
\right)- \left(
\frac{dn_e}{dr} \Delta \mu_e +\frac{dn_b}{dr} \delta \mu^{\infty}_n \right) {\pmb e}_r
=\frac{1}{4\pi} \, [{\pmb \nabla}\times {\pmb H}_{\rm c1}]\times {\pmb B},
\label{fb2}
\end{align}
where ${\pmb H}_{\rm c1}$ is the vector directed along ${\pmb B}$, 
whose absolute value 
equals the lower critical magnetic field for a simplified model 
of non-interacting proton vortices \cite{gas11,gd16}.%
%
\footnote{We assume that protons form type-II superconductor.
Note that in the superconducting $npe$-matter chemical potentials (and other thermodynamic quantities) 
depend not only on $n_b$, $n_e$, and $T$, but also on the magnetic field ${\pmb B}$ \cite{gas11,gd16}.}
%
This equation can be easily solved \cite{gl16} 
for $\Delta \mu_e$ and $\delta \mu_n^\infty$, 
similar to how it is done in Sec.\ \ref{perturb},
so that all other thermodynamic quantities can be determined.
The remaining scheme of the solution is also quite similar.
However, the problem is slightly
more delicate than before
since now the magnetic field is confined to flux tubes (proton vortices)
and one should 
accurately account for \underline{both} ordinary diffusion of ``nonsuperfluid'' particles,
as well as various dissipative (and non-dissipative) processes associated 
with particle interaction with the flux tubes.
The complex dynamic equations describing these effects 
have been (partly) formulated in Refs.\ \cite{gas11, gd16}; 
full account is given in Ref.\ \cite{gd17}.
Application of these equations to the problem considered here 
is a subject of future work.

\subsection{Accounting for deviations from the diffusion and beta-equilibrium, which are not related to the magnetic field}
\label{deviations}

In Sec.\ \ref{perturb} we assumed that a deviation of the star from the diffusion
and beta-equilibrium is exclusively determined by the magnetic field. 
This assumption allowed us to neglect the terms $\partial n_\alpha/\partial t$
in the continuity equations (\ref{cont}).
But how our scheme will be modified if
some part of the 
deviation from the diffusion and beta-equilibrium
is not related to the magnetic field?
For example, additional deviation can arise due to compression 
of the spinning down neutron star
or simply due to its cooling 
(if one accounts for a weak dependence of chemical potentials on $T$).
In this situation one should start with the most general form
of the continuity equations 
[cf.\ Eqs.\ (\ref{cont2p2}) and (\ref{cont2n2})], 
\begin{align}
& \frac{\partial n_p}{\partial t} 
+ n_p \, {\rm div} \,{\pmb U} + \frac{d n_p}{dr} \, U_r =-{\rm div} \,{\pmb \nu}_{p}-\Delta \Gamma,&
\label{cont2p2full}\\
& \frac{\partial n_n}{\partial t} 
+n_n \, {\rm div} \, {\pmb U} + \frac{d n_n}{dr} \, U_r =-{\rm div} \,{\pmb \nu}_{n}+\Delta \Gamma.&
\label{cont2n2full}
\end{align}

To simplify presentation, 
below we assume
that, initially, 
there is a 
deviation from the diffusion and beta-equilibrium,
which is not caused exclusively by 
the magnetic field, 
but the subsequent evolution of the system 
proceeds with the magnetic field as the only perturbing factor.
Then the system should evolve to the configuration 
studied in detail in Sec.\ \ref{perturb} 
on some typical timescale $\tau_{0}$, 
which is, as a rule, much smaller than the typical magnetic timescale $\tau_B$ %
%
\footnote{It can be shown that the typical timescale for reaching the diffusion equilibrium 
in this problem is
$\tau_0 \sim R^2 J_{np}/(\mu_n n_b)$ and for reaching the beta-equilibrium is 
$\tau_0 \sim n_p^2/(\mu_n n_b \lambda_e)$ (see Sec.\ \ref{dissip} for the definition of $\lambda_e$). }.
%
Generalization of our approach to the case when 
some other factors (besides the magnetic field)
perturb the system out of the diffusion and beta-equilibrium
during its evolution (e.g., decreasing temperature)
is rather straightforward and can be made in a similar fashion.

The partial derivatives 
$\partial n_\alpha/\partial t$ ($\alpha=n$, $p$)
in Eqs.\ (\ref{cont2p2full}) and (\ref{cont2n2full})
can be expressed through $\delta \mu_n^\infty$ and $\Delta \mu_e$ as
\begin{align}
\frac{\partial n_\alpha }{\partial t} = \frac{\partial n_\alpha}{\partial \mu_n} 
\, \frac{\partial  \mu_n}{\partial t}+
\frac{\partial n_\alpha}{\partial \Delta \mu_e} 
\, \frac{\partial  \Delta \mu_e}{\partial t}=
\frac{\partial n_\alpha}{\partial \mu_n} 
\, \frac{\partial  \delta \mu_n^\infty}{\partial t}+
\frac{\partial n_\alpha}{\partial \Delta \mu_e} 
\, \frac{\partial  \Delta \mu_e}{\partial t},
\label{x1}
\end{align}
where we, for simplicity, 
presented $n_{n}$ and $n_p$ as functions of only
$\mu_n$ and $\Delta \mu_e$ 
(thus assuming that the dependence of $n_\alpha$ on $T$ can be neglected)
and used the Cowling approximation, $\delta \mu_n=\delta \mu_n^\infty$ 
[cf.\ Eq.\ (\ref{dm2})].
To calculate the time derivatives in the right-hand side of Eq.\ (\ref{x1}) one should use 
the expression (\ref{solve}) for $\delta \mu_n^\infty$ and $\Delta \mu_e$.
As a result, one will obtain two types of terms.
The terms of the first type depend on $\partial {\pmb B}/\partial t$, 
hence their typical 
timescale is $\tau_{\rm B}$ and they 
drop out from 
the continuity equations (\ref{cont2p2full}) and (\ref{cont2n2full}) 
to leading order in $B^2/P$ 
because of the very same reasons that
have already been discussed in the beginning 
of Sec.\ \ref{perturb2}.
The terms of the second kind depend on $\partial \zeta(r,\,t)/\partial t$
and cannot a priori be neglected when there is an initial disturbance in
the system, which is not related to the magnetic field.
Therefore, one should substitute $\partial n_\alpha/\partial t$ 
into Eqs.\ (\ref{cont2p2full}) and (\ref{cont2n2full})
in the form [see Eqs.\ (\ref{Z11})--(\ref{solve})]
\begin{align}
&\frac{\partial n_\alpha}{\partial t}
=
\left(\frac{\partial n_\alpha}{\partial \mu_n} \,\,\,\, \frac{\partial n_\alpha}{\partial \Delta \mu_e} \right)
\left(
\begin{array}{cc}
\,\,\,\, n_e & \,\,\,\,n_b\\
-\frac{dn_e}{dr} & -\frac{dn_b}{dr}
\end{array}
\right)^{-1}
\left(
\begin{array}{c}
\,\,\partial \zeta(r,\, t)/\partial t \\
-\partial \zeta'(r,\, t)/\partial t
\end{array}
\right).&
\label{solve0}
\end{align}
Equations (\ref{cont2p2full}) and (\ref{cont2n2full})
can then be solved for ${\rm div} \,  {\pmb U}$ and $U_r$,
which allows one to determine 
$U_\theta$ 
from
Eq.\ (\ref{Uth}) with $\xi=0$.
The main difference from the results of the previous sections
is that now $U_r$ and $U_\theta$ 
depend not only on $\zeta$ and its spatial derivatives,
but also on $\partial \zeta/\partial t$.
An equation for $\zeta(r,t)$ 
can be obtained in the same way as in Sec.\ \ref{perturb44a} and
is given by the condition (\ref{Uthetacond}).
However, now it is a {\it partial} differential equation; 
it should thus be supplemented by the initial condition, 
$\zeta(r,\,0)$, and
by the boundary conditions, 
following, in particular, from the regularity of $U_r$ at $r \rightarrow 0$.

\section{Magnetic field dissipation}
\label{dissip}

The aim of the present section is to derive
a general expression for the total dissipation rate $\dot{E}_{\rm B}$ 
of the magnetic field energy for the system in the quasistationary state,
free of any specific approximations. 
In what follows, all the surface integrals appearing in the formulas are ignored
for simplicity; 
they can be easily written out if necessary.
One has
\begin{align}
\dot{E}_{\rm B}= \frac{1}{4\pi} 
\int_V {\pmb B}\frac{\partial{\pmb B}}{\partial t}\, dV.
\label{Ebdot1}
\end{align}
This equation can be represented as (e.g., Ref.\ \cite{gr92})
\begin{align}
\dot{E}_{\rm B}=-\int_V \pmb{Ej}\, dV.
\label{EBdiss}
\end{align}

Let us express the electric field, entering Eq.\ (\ref{EBdiss}), 
from Eq.\ (\ref{33x}) for protons ($\alpha=p$) 
with the vanishing left-hand side,
\begin{eqnarray}
{\pmb E}=- \frac{{\pmb u}_p\times {\pmb B}}{c}
+\frac{{\pmb \nabla} \mu_p^\infty}{e}
+\frac{ J_{ep} ({\pmb u}_p-{\pmb u}_e) + J_{np} ({\pmb u}_p - {\pmb u}_n)
}{ e n_e}, 
\label{EE}
\end{eqnarray}
where $e \equiv e_p$ and we make use of the quasineutrality conditon, $n_e=n_p$,
and the definition of $\mu_p^\infty$ from Sec.\ \ref{perturb2}.
The second term in Eq.\ (\ref{EE}) is potential and thus does not contribute 
to the magnetic field dissipation 
(see, e.g., Ref.\ \cite{kg17} for more details). 
Thus,
\begin{eqnarray}
\dot{E}_{\rm B}=-\int_V \left[- \frac{{\pmb u}_p\times {\pmb B}}{c}
+\frac{ J_{ep} ({\pmb u}_p-{\pmb u}_e) + J_{np} ({\pmb u}_p - {\pmb u}_n)}{ e n_e}
\right]\pmb{j}\, dV,
\label{dotE2}
\end{eqnarray}
The first term here can be modified:
\begin{eqnarray}
\int_V \left(\frac{{\pmb u}_p\times {\pmb B}}{c}\right)
\pmb{j}\, dV=-\int_V \left(\frac{{\pmb j}\times {\pmb B}}{c}\right)\pmb{u}_p\, dV.
\end{eqnarray}
Substituting now Eqs.\ (\ref{4444xx}) and (\ref{dm1}), we obtain
\begin{eqnarray}
-\int_V \left(\frac{{\pmb j}\times {\pmb B}}{c}\right)\pmb{u}_p \, dV
=\int_V \left(-n_e {\pmb \nabla} \Delta \mu_e 
-n_b {\pmb \nabla} \delta \mu_n^\infty \right)\pmb{u}_p\, dV.
\label{Wx0}
\end{eqnarray}
Integration by parts of the first term in the right-hand side of this equation gives 
(we remind that we skip the surface integral)
\begin{eqnarray}
\int_V \left[{\rm div} (n_p \pmb{u}_p) \Delta \mu_e
- {\pmb \nabla} \delta \mu_n^\infty(n_n {\pmb u}_n + n_p {\pmb u}_p) 
- {\pmb \nabla} \delta \mu_n^\infty \, n_n ({\pmb u}_p - {\pmb u}_n)\right]\, dV,
\label{Wx}
\end{eqnarray}
where we 
expressed $\pmb{u}_p$ in the second term as 
$\pmb{u}_p=\frac{n_n {\pmb u}_n + n_p {\pmb u}_p}{n_b} 
+ \frac{n_n ({\pmb u}_p - {\pmb u}_n)}{n_b}$. 
Now, 
(i) to transform the first term we make use 
of the proton continuity equation, 
${\rm div} (n_p \pmb{u}_p)
=-\Delta \Gamma$;
(ii) to transform the second term 
we integrate it by parts and use the baryon continuity equation, 
${\rm div} (n_n \pmb{u}_n+n_p \pmb{u}_p)=0$; 
(iii) to transform the third term we express ${\pmb \nabla} \delta \mu_n^\infty$ 
from Eq.\ (\ref{33x}) for neutrons, 
which reads
\begin{align}
n_n {\pmb \nabla} \delta\mu_n^\infty = -\sum_{\beta\neq n} J_{n\beta}({\pmb u}_n-{\pmb u}_\beta).
\label{neutron}
\end{align}
As a result, we get
\begin{eqnarray}
\int_V \left[
-\Delta \mu_e \Delta \Gamma
- J_{en} ({\pmb u}_e - {\pmb u}_n)({\pmb u}_p - {\pmb u}_n) 
- J_{np} ({\pmb u}_p - {\pmb u}_n)^2  \right]\, dV.
\end{eqnarray}
Returning then to Eq.\ (\ref{dotE2}) and rearranging terms, 
we obtain
\begin{eqnarray}
\dot{E}_{\rm B}=-\int_V \pmb{Ej}\, dV=
\int_V \left[
-\Delta \mu_e \Delta \Gamma
-J_{en} ({\pmb u}_e - {\pmb u}_n)^2 
- J_{ep} ({\pmb u}_e - {\pmb u}_p)^2 - J_{np} ({\pmb u}_n - {\pmb u}_p)^2 \right]\, dV \nonumber \\
+\int_V ({\pmb u}_e - {\pmb u}_p) \left[J_{en} ({\pmb u}_e - {\pmb u}_n) 
+ J_{np} ({\pmb u}_p - {\pmb u}_n)\right] \, dV.
\label{Wx2}
\end{eqnarray}
Let us show that the last term in the right-hand side of Eq.\ (\ref{Wx2}) vanishes.
Using Eq.\ (\ref{neutron}), one may write
\begin{eqnarray}
\int_V ({\pmb u}_e - {\pmb u}_p) 
\left[J_{en} ({\pmb u}_e - {\pmb u}_n) + J_{np} ({\pmb u}_p - {\pmb u}_n)\right] \, dV
=\int_V ({\pmb u}_e - {\pmb u}_p) n_n {\pmb \nabla} \delta \mu_n^\infty \, dV= 
\nonumber \\
\int_V ({\pmb u}_e - {\pmb u}_p) n_b {\pmb \nabla} \delta \mu_n^\infty \, dV
-\int_V ({\pmb u}_e - {\pmb u}_p) n_e {\pmb \nabla} \delta \mu_n^\infty \, dV.
\label{Wx3}
\end{eqnarray}
Equation (\ref{4444xx}) implies that 
$({\pmb u}_e-{\pmb u}_p) n_b {\pmb \nabla} \delta \mu_n^\infty 
=-({\pmb u}_e-{\pmb u}_p)n_e {\pmb \nabla} \Delta \mu_e^\infty$. 
Using this equality together with the 
charge conservation equation, ${\rm div}\, {\pmb j}=0$,
and 
integrating  by parts both terms in the right-hand side of Eq.\ (\ref{Wx3}), 
one verifies that Eq.\ (\ref{Wx3}) indeed vanishes.
Consequently,
\begin{eqnarray}
\dot{E}_{\rm B}=
\int_V \left[-\Delta \mu_e \Delta \Gamma
-J_{en} ({\pmb u}_e - {\pmb u}_n)^2 - J_{ep} ({\pmb u}_e - {\pmb u}_p)^2 
- J_{np} ({\pmb u}_n - {\pmb u}_p)^2 \right]\, dV. 
\label{EBdissnpe}
\end{eqnarray}
We see that the magnetic field dissipates because of 
particle mutual transformations and relative motion (diffusion). 
If we neglect (weak) interaction between electrons and neutrons, 
i.e. put $J_{en}=0$, then $\dot{E}_{\rm B}$ will take the familiar form 
(see, e.g., Ref.\ \cite{gr92}),
\begin{eqnarray}
\dot{E}_{\rm B}=
\int_V \left[-\Delta \mu_e \Delta \Gamma
 - \frac{j^2}{\sigma_0} - J_{np} ({\pmb u}_n - {\pmb u}_p)^2 \right]\, dV, 
\label{simpl}
\end{eqnarray}
where $\sigma_0=e^2 n_e^2/J_{ep}$ is the electrical conductivity 
in the absence of the magnetic field.
The last term in the right-hand side of Eq.\ (\ref{simpl})
describes the effect of ambipolar diffusion.
The associated ambipolar velocity, 
${\pmb u}_{\rm p}-{\pmb u}_n$,
can be expressed through ${\pmb \nabla} \mu_n^\infty$ from Eq.\ (\ref{neutron}).
In contrast to the results of Refs.\ \cite{gr92,td96,hrv08,reisenegger09,hrv10,gjs11,bl16,papm17}, both quantities
$\Delta \mu_e$ and ${\pmb \nabla} \mu_n^\infty$ 
are almost independent of 
the relaxation time $\tau_{np}$
and beta-reaction rate.%
%
\footnote{To make this statement more precise, see Appendix \ref{A3}.}
%
As a consequence, the ambipolar diffusion timescale can be estimated as 
(see Sec.\ \ref{example} for more details):
$\tau_{\rm B} \sim B^2/[J_{np} ({\pmb u}_p-{\pmb u}_n)^2]
\sim n_p m_p R^2/(B^2 \tau_{pn})$.
This estimate coincides 
with the {\it solenoidal} ambipolar diffusion timescale introduced in Ref.\ \cite{gr92} 
(see Eq.\ (34) there).
Note that the {\it irrotational} diffusion timescale of Ref.\ \cite{gr92} 
(see also Refs.\ \cite{td96,act04,hrv08,reisenegger09,dss09,hrv10,gjs11,dgp12,bl16,papm17})
does not appear in our analysis.

Proceeding in a very similar way in the case of $npe\mu$-matter, we obtain
\begin{align}
&\dot{E}_{\rm B}=
\int_V \left[-\Delta \mu_e \Delta \Gamma -\Delta \mu_\mu \Delta \breve{\Gamma}
-\frac{1}{2}\sum_{\alpha, \beta=n,\, p, \, e, \, \mu} 
J_{\alpha\beta}({\pmb u}_\alpha - {\pmb u}_\beta)^2
\right]\, dV,& 
\label{EBdissnpemu}
\end{align}
where the source $\Delta \breve{\Gamma}$ is introduced in Sec.\ \ref{perturb5}.
If we are in {\it subthermal} regime, i.e., 
$\Delta \mu_\mu/(k_{\rm B}T) \ll 1$ [or $\Delta \mu_e/(k_{\rm B}T) \ll 1$] then
$\Delta \breve{\Gamma}$ (or $\Delta \Gamma$) 
can be approximately presented as 
$\Delta \breve{\Gamma} = \lambda_\mu \Delta \mu_\mu$
(or $\Delta \Gamma = \lambda_e \Delta \mu_e$),
where $\lambda_\mu>0$ and $\lambda_e>0$ are temperature- and density-dependent 
beta-reaction coefficients given in, e.g., Ref.~\cite{ykgh01}.

In Ref.\ \cite{kg17} it is shown that Eqs.\ (\ref{EBdissnpe}) and (\ref{EBdissnpemu})
retain its form in the case of superfluid matter.
Eqs.\ (\ref{EBdissnpe}) and (\ref{EBdissnpemu}) have a clear physical interpretation.
It can be demonstrated 
that the right-hand sides of these equations equal to the (minus) entropy generation rate $\dot{S}$
[excluding the thermal conductivity and thermo-diffusion contributions, 
which were neglected
in the dynamic equations of Sec.\ \ref{maineq}].
In fact, this result is a special case 
of a more general theorem, 
which can be formulated as follows.

\vspace{0.2 cm}
\noindent

{\bf Theorem:} Assume that the system is quasistationary 
in a sense described in Sec.\ \ref{perturb2}.
Then the rate of change of the magnetic field energy $\dot{E}_B$ in the volume $V$
is given by
\begin{align}
\dot{E}_{\rm B} = -\int_V T \dot{S} \,dV +{\rm ``Surface \, terms"},
\label{theo}
\end{align}
where the first term is the \underline{total} heat generated in the system 
($\dot{S}$ is the rate of change of the entropy density)
and the second term represents possible magnetic energy 
and/or particle (e.g., neutrino)
flows 
through the boundary of the volume $V$.
This theorem should work equally well for both normal and 
superfluid/superconducting magnetized matter (in the latter case $\dot{E}_{\rm B}$
is the total vortex energy, including their kinetic energy).
It is non-trivial, since it forbids, in particular, 
transformation of $E_{\rm B}$
into the energy of macroscopic flows or into the ``chemical'' energy 
(when $\Delta \mu_e$ increases).
The proof will be presented elsewhere.

\vspace{0.2 cm}
\noindent

Note that the dissipation rate $\dot{E}_{\rm B}$ calculated above depends on the 
differences ${\pmb u}_\alpha-{\pmb u}_\beta={\pmb \nu}_\alpha/n_\alpha-{\pmb \nu}_\beta/n_\beta$
[see Eq.\ (\ref{coord2})]. 
The vectors ${\pmb \nu}_\alpha$ are, in turn, expressed through various 
chemical potentials and the magnetic field by the formula (\ref{nui}).
Thus, $\dot{E}_{\rm B}$ can be calculated
(even without knowing the velocity ${\pmb U}$),
provided that these chemical potentials are determined. 
The next section presents an example of such calculation.

\vspace{0.2 cm}
\noindent

 {\bf Remark.}
The theorem (\ref{theo}) is valid as long as one can neglect 
the time derivatives in the continuity equations (\ref{cont2p2}) and (\ref{cont2n2}).
This is not the case if there are some other factors (except for the magnetic field)
that disturb the system from the diffusion and beta-equilibrium 
(see Sec.\ \ref{deviations} for an example of such situation).
Then $\partial n_\alpha/\partial t$ can not generally be neglected
and Eq.\ (\ref{theo}) should be replaced with 
\begin{align}
\dot{E}_{\rm B} = -\int_V T \dot{S} \,dV 
-\int_V \delta \mu_n^\infty \, \frac{\partial n_b}{\partial t} \, dV
-\int_V \Delta \mu_e \, \frac{\partial n_p}{\partial t}\, dV
+{\rm ``Surface \, terms"},
\label{theo2}
\end{align}
where the last two integrals can be evaluated by making use of 
Eq.\ (\ref{solve}) and expressions for $\partial n_\alpha/ \partial t$ ($\alpha=n$, $p$).
For an illustrative example of Sec.\ \ref{deviations}
$\partial n_\alpha/ \partial t$ is given by Eq.\ (\ref{solve0}).

\section{Numerical example}
\label{example}

For illustration, here we present detailed calculations 
of the magnetic field dissipation rate $\dot{E}_{\rm B}$ 
for normal $npe$-matter using the formula (\ref{simpl}) 
[i.e., assuming $J_{en}=0$].%
%
\footnote{In fact, this simple example admits also relatively straightforward calculation 
of the components $U_r$ and $U_\theta$ of the flow velocity ${\pmb U}$ 
(see Appendix \ref{A3}, 
where the components $u_{nr}$ and $u_{n\theta}$ 
of the neutron velocity ${\pmb u}_n$ are calculated). 
We, however, plan to find all the components of ${\pmb U}$ in a future work.}
%
Then, using Eqs.\ (\ref{rotB}), (\ref{neutron}) 
and $\sigma_0 = e^2 n_e^2/J_{ep}$, one can rewrite Eq.\ (\ref{simpl}) as
\begin{align}
\dot{E}_{\rm B}=
- \int_V \left[\Delta \mu_e \Delta\Gamma
+ \left( \frac{c}{4\pi e n_e} \right)^2 J_{ep} \left( \mathrm{rot}\boldsymbol{B} \right)^2 + \frac{n_n^2}{J_{np}}\, \left({\pmb \nabla}\delta \mu_n^\infty \right)^2 \right]
\, dV.
\label{simpl2}
\end{align}
In what follows, we take $J_{ep}$ and $J_{np}$ from Refs.~\cite{ys90, ys91b};
$\Delta\Gamma$ due to non-equilibrium modified Urca (hereafter MUrca) processes 
[denoted as $\Delta\Gamma^\text{({\rm MU})}$] 
is taken in the same simple form as in Ref.~\cite{papm17} 
(see also references therein), 
but with the non-linear corrections from Refs.\ \cite{reisenegger95, ykgh01};
for $\Delta\Gamma^\text{(DU)}$ due to non-equilibrium direct Urca (hereafter DUrca) 
process we employ the exact expression listed in Refs.\ \cite{reisenegger95,ykgh01}, 
but set the effective masses of nucleons to
$m_p^\ast = 0.7m_p$, $m_n^\ast = 0.7m_n$:
\begin{align}
J_{ep} &\approx 2.0\times 10^{28} \, T_8^2 \left( \frac{\rho_0}{\rho} \right)^{5/3} \left( \frac{n_e}{n_0} \right)^{4/3} \frac{\text{g}}{\text{cm}^3\,\text{s}}, 
\label{sigma}\\
J_{np} &\approx 1.25\times 10^{31} \, T_8^2 \left( \frac{\rho_0}{\rho} \right)^{1/3} 
\left( \frac{n_p}{n_0} \right) \,
 \frac{\text{g}}{\text{cm}^3\,\text{s}}, 
\label{Jnp}\\
\Delta\Gamma^\text{(MU)} &\approx 5\times 10^{27} \, \frac{\Delta\mu_e}{\text{erg cm$^3$ s}} T_8^6 \left( \frac{\rho}{\rho_0} \right)^{2/3}\left[ 1 + \frac{189}{367}\left( \frac{\Delta\mu_e}{\pi k_B T} \right)^2 + \frac{21}{367}\left( \frac{\Delta\mu_e}{\pi k_B T} \right)^4 + \frac{3}{1835}\left( \frac{\Delta\mu_e}{\pi k_B T} \right)^6 \right], 
\label{lambdaM}\\
\Delta\Gamma^\text{(DU)} &\approx 1.6\times 10^{36} \, \frac{\Delta\mu_e}{\text{erg cm$^3$ s}} T_8^4 \left( \frac{n_e}{n_0} \right)^{1/3} \left[ 1 + \frac{10}{17}\left( \frac{\Delta\mu_e}{\pi k_B T} \right)^2 + \frac{1}{17}\left( \frac{\Delta\mu_e}{\pi k_B T} \right)^4 \right]. 
\label{lambdaD}
\end{align}
Here $\rho=\varepsilon/c^2$ is the density;
$\rho_0 = 2.8\times 10^{14}$~g~cm$^{-3}$~s$^{-1}$;
$n_0 = 0.16$~fm$^{-3}$; 
$T_8 = T/(10^8\,\text{K})$. 
The first three equations (\ref{sigma})--(\ref{lambdaM}) 
are based on a rather outdated microphysics
and are used here for simplicity. 
We checked, however, that more accurate (but lengthy) expressions for 
$J_{ep}$
and $\Delta \Gamma^{({\rm MU})}$, available in the literature 
(see, e.g., Refs.\ \cite{shternin08,ykgh01}), 
do not affect our results much.
Note that, in Eqs.\ (\ref{lambdaM}) and (\ref{lambdaD}) 
we employ the non-linear expressions 
for $\Delta \Gamma^{({\rm MU})}$ and $\Delta \Gamma^{({\rm DU})}$
valid at arbitrary ratio of $\Delta \mu_e/(k_{\rm B}T)$ 
(not only at $\Delta \mu_e \ll k_{\rm B}T$).

Using Eqs.\ (\ref{neutron}), (\ref{Jnp}), and the results of Appendix \ref{A1},
it is straightforward to estimate the typical difference between the neutron and proton velocities,
$|{\pmb u}_n-{\pmb u}_p| 
\sim \widetilde{B}^2/(4 \pi \, J_{np} R)
\sim 3 \times 10^{-10} \, \widetilde{B}_{14}^2/(T_8^2 R_6)$~cm~s$^{-1}$,
where $\widetilde{B}_{14}$ is a typical magnetic field in units of $10^{14}$~G
and $R_6$ is a typical lengthscale in units of $10^6$~cm.
This result should be compared with an estimate for $|{\pmb u}_e-{\pmb u}_p|$,
following from Eqs.\ (\ref{rotB}) and (\ref{j}):
$|{\pmb u}_e-{\pmb u}_p| \sim  B c/(4\pi \, e n_e R) 
\sim 10^{-11} \, \widetilde{B}_{14}/R_6$~cm~s$^{-1}$.

To evaluate the integral (\ref{simpl2}) 
we need to specify the magnetic field and then, 
using it, determine the 
functions $\Delta \mu_e$ and $\delta \mu_n^\infty$ 
from the formulas given in Appendix \ref{A1}.
For numerical calculations, 
we choose the toroidal-poloidal magnetic field configuration 
from Ref.~\cite{papm17} (see Sec.~3 there and our Appendix~\ref{A2}). 
We adopt the three models of the magnetic field, 
which differ by the ratio of maximum absolute 
values of toroidal and poloidal fields, 
$B_{\rm T max}/B_{\rm P max}$ (see Tab.~\ref{theTable}).
The first and the last of these models coincide with, respectively,
the models A and B from Ref.\ \cite{papm17}.

We also need the (equilibrium) radial profiles of the functions 
$\rho(r)$, $n_b(r)$, and $n_e(r)$ in the stellar core.
To calculate them we employed HHJ equation of state \cite{hhj99}, 
which gives the circumferential radius $R_{\rm NS} = 12.2$~km 
for a model of an NS with the mass $M=1.4\,M_\odot$.
Note that DUrca process is forbidden for a chosen NS model.
However, to get an impression of a possible effect 
of non-equilibrium processes which are
stronger than MUrca, 
we artificially switched DUrca on in one of our models (in the whole core).%
%
\footnote{One should bear in mind that even if DUrca is closed, 
there could be other very powerful non-equilibrium processes 
of particle mutual transformations
if we allow for hyperons in the NS core \cite{jones01,lo02,hly02}.
To our knowledge, these non-leptonic processes were ignored in the literature
devoted to the magnetic field evolution,
but they can be very effective dissipation agents.}
%

Using these models and the formulas from Appendix~\ref{A1},
we calculate the functions $\Delta \mu_e$ and $\delta \mu_n^\infty$
[to do this, we also need to know the function $\zeta(r)$, see Eqs.\ (\ref{Z11}) and (\ref{Z22});
it is calculated in Appendix \ref{A3} 
following the general procedure described in Sec.\ \ref{perturb44a}]. 
Then we have all the necessary information 
to calculate the integral (\ref{simpl2}). 
Choosing 
$\widetilde{B} = \max \left\{ B_{\rm P max}, 
B_{\rm T max} \right\}$ 
and $\widetilde{n} = n_0$ 
in Eqs.~(\ref{dmue}) and (\ref{dmun}), 
and integrating over the whole NS core, we find
\begin{equation}
\dot{E}_{\rm B} = - \alpha_\text{R}^{({\rm type})} \widetilde{B}_{14}^4 T_8^k \left[ 1 + \beta_2^{({\rm type})} \frac{\widetilde{B}_{14}^4}{T_8^2} + \beta_4^{({\rm type})} \frac{\widetilde{B}_{14}^8}{T_8^4} + \beta_6^{({\rm type})} \frac{\widetilde{B}_{14}^{12}}{T_8^6} \right] - \alpha_\text{Ohm} \widetilde{B}_{14}^2 T_8^2 - \alpha_\text{Amb} \frac{\widetilde{B}_{14}^4}{T_8^2},
\label{Edot_final}
\end{equation}
where $k=6$ for MUrca (type = MU) and $k=4$ for DUrca (type = DU) processes, 
and the coefficients $\alpha$ and $\beta$ are listed in Table~\ref{theTable}. 
%
\begin{table}
	\begin{center}
		\caption{\label{theTable} Numerical coefficients in Eqs.~(\ref{Edot_final}) and (\ref{Etotal}) for 
			an NS with $M=1.4 M_\odot$. 
			Abbreviation `MU' and `DU' stands for MUrca  and DUrca processes 
			as the main neutrino emission mechanisms, respectively.}
		\renewcommand{\arraystretch}{1.4}
		\begin{tabular}{cccccccccccc}
			\hline\hline
			$B_{\rm T max}/B_{\rm P max}$ & $\gamma$ [$10^{44}$ erg] & \multicolumn{2}{c}{$\alpha_\text{R}$ [$10^{23}$ erg/s]} & \multicolumn{2}{c}{$\beta_2\times 10^8$} & \multicolumn{2}{c}{$\beta_4\times 10^{15}$} & \multicolumn{2}{c}{$\beta_6\times 10^{24}$} & $\alpha_\text{Ohm}$ [$10^{25}$ erg/s] & $\alpha_\text{Amb}$ [$10^{30}$ erg/s] \\
			& & $\;\;$MU$\;$ & $\;$DU$\;\;$ & $\;\;$MU$\;$ & $\;$DU$\;\;$ & $\;\;$MU$\;$ & $\;\;$DU$\;\;$ & $\;$MU$\;$ & $\;$DU$\;\;$ & & \\
			\hline
			0    & $4.0$ & $1.1$ & $1.3\times 10^{8}$ & $1.0$  & $1.3$  & $0.09$  & $0.1$   & $0.3$  & 0 & $0.52$ & $1.1$ \\
			1    & $5.1$ & $6.5$ & $7.5\times 10^{8}$ & $6.0$  & $6.9$  & $1.3$   & $1.3$   & $8.2$  & 0 & $3.7$  & $9.8$ \\
			2.29 & $1.9$ & $5.1$ & $5.9\times 10^{8}$ & $6.5$  & $7.5$  & $1.4$   & $1.4$   & $8.0$  & 0 & $3.3$  & $7.9$ \\
			\hline
		\end{tabular}
	\end{center}
\end{table}
%
\begin{figure}
	\includegraphics[width=\textwidth]{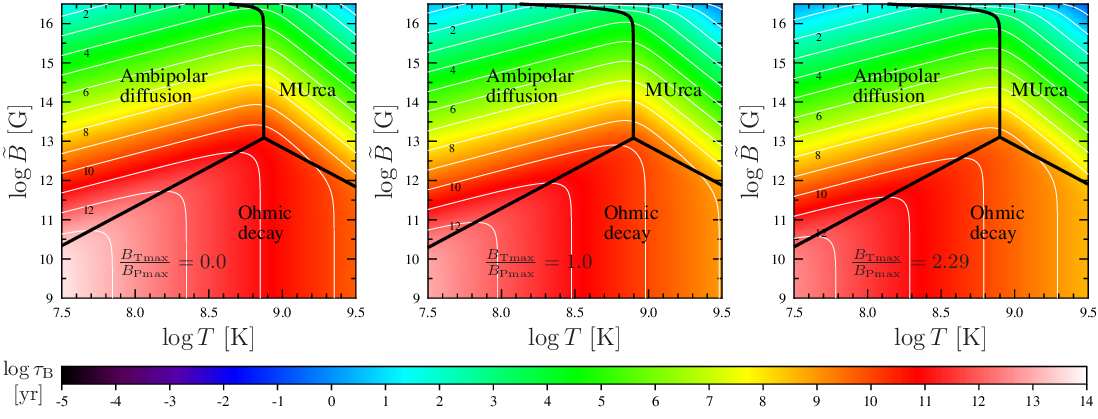}
	\caption{\label{fig:MU} The magnetic field decay timescale 
$\tau_{\rm B} = 2 E_{\rm B}/ \dot{E}_{\rm B}$ in the case of the non-equilibrium MUrca processes 
as the main mechanism restoring chemical equilibrium. 
From left to right: $B_{\rm T max}/B_{\rm Pmax} = 0,\, 1,\, 2.29$. 
Thin white lines correspond to $\log \tau_{\rm B} = {\rm const}$. 
Thick black lines in the ($\log\widetilde{B} - \log T$) plane
separate
the regions where one of the three dissipation mechanisms 
(ambipolar diffusion, MUrca processes, or Ohmic decay) is most efficient.}
\end{figure}
%
\begin{figure}
	\includegraphics[width=\textwidth]{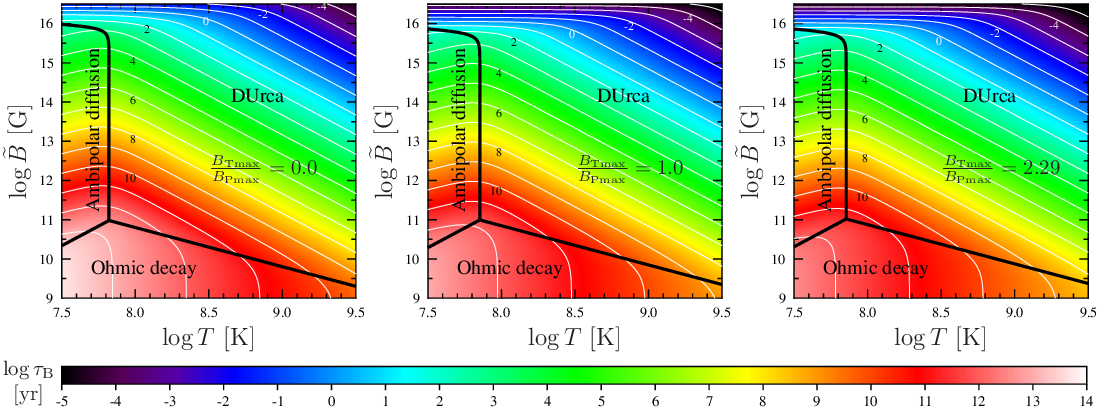}
	\caption{\label{fig:DU} The same as in Fig.~\ref{fig:MU} but for the non-equilibrium DUrca process as the main mechanism that restores chemical equilibrium.}
\end{figure}
%
One can compare this dissipation rate with the total magnetic field energy stored in the core,
\begin{equation}
E_{\rm B} = \int_V \frac{\boldsymbol{B}^2}{8\pi} dV = \gamma \widetilde{B}_{14}^2,
\label{Etotal}
\end{equation}
where the numerical factor $\gamma$ is also listed in Table~\ref{theTable}. 
Figures~\ref{fig:MU} and \ref{fig:DU} 
display the characteristic magnetic field decay timescale, 
$\tau_{\rm B} \equiv 2 E_{\rm B}/\dot{E}_{\rm B}$, 
due to the mechanisms described above. 
Thick black lines separate regions where the contribution into $\tau_{\rm B}$ 
of one or another term in Eq.~(\ref{Edot_final})  is dominant.
Thus, in the `MUrca' and `DUrca' domains 
the non-equilibrium beta-processes are the most important 
[first term in Eq.\ (\ref{Edot_final})];
in the ``Ohmic decay'' domain (second term)
ohmic dissipation prevails;
finally in the ``ambipolar diffusion'' domain 
the third term 
mostly determines the timescale~$\tau_{\rm B}$.
As we have already emphasized in Sec.\ \ref{dissip},
this timescale coinci\begin{flushright}
	
\end{flushright}des with 
the {\it solenoidal} ambipolar timescale from Ref.\ \cite{gr92}.

As follows from the analysis of the figures,
the boundary between the ambipolar diffusion and reaction (MUrca or DUrca) domains 
is independent of $\widetilde{B}$ at $\widetilde{B}\lesssim 10^{15}$~G. 
This means that the non-linear terms [see square bracket in Eq.~(\ref{Edot_final})] 
are not important 
at such ${\widetilde B}$ and 
can be neglected. 
The non-linear regime of beta-reactions rapidly switches on 
at $\widetilde{B} \gtrsim 10^{16}$~G (MUrca) or $3\times 10^{15}$~G (DUrca). 
Note, however, that the magnetic fields that large become quantizing,
which may affect the results quantitatively 
(but not qualitatively,
see, e.g., Refs.\ \cite{ys91a, by99} and figure 3 in Ref.\ \cite{ys91b}).
Second, there is a clear separation between the domains:
the non-equilibrium beta-processes prevail at high $\widetilde{B}$ and $T$;
ambipolar diffusion becomes important at relatively low temperature
(the corresponding timescale scales as $1/T^2$), 
while the Ohmic decay plays a dominant role at low magnetic fields,
but then the typical timescale exceeds the age of the Universe.
Finally, one may note that when DUrca is switched on, it becomes
the main dissipation mechanism 
in the almost whole region of $\widetilde{B}$ and $T$ shown in Fig.\ \ref{fig:DU}.
Moreover, a typical timescale $\tau_{\rm B}$ for this mechanism
can be very small, about a century for $T\gtrsim 6\times 10^8$~K and $\tilde{B}\sim 10^{14}$~G 
and 4--7 days for $\tilde{B}\sim 10^{16}$~G. 
The magnetic field will reconfigure (by effective dissipation) on these short timescales
in order to vanish $\Delta \mu_e$ in the core, provided that the system evolves in the 
subthermal regime ($\Delta \mu_e \lesssim k_{\rm B}T$).
The case of the suprathermal regime ($\Delta \mu_e \gtrsim k_{\rm B}T$) 
is a bit more tricky and will be analysed by us elsewhere.

\section{Conclusions and final remarks}
\label{concl}

In this work we study the quasistationary equilibrium and dissipation 
in magnetized cores of NSs.
We argue that the generally accepted approach to this problem 
pioneered by Goldreich and Reisenegger \cite{gr92} 
(see also Refs.~\cite{td96,hrv08,reisenegger09,hrv10,gjs11,bl16,papm17})
should be revised (see Appendix \ref{critique} for details). 
Taking, as an example, normal $npe$-matter in NS cores, 
we formulate a general scheme allowing one to find
all the necessary ingredients 
(thermodynamic parameters, velocities, electric field, etc.)
to self-consistently follow the quasistationary evolution of the stellar magnetic field.
Our results can be summarized as follows:

\begin{itemize}
	
\item 
Expanding the quantities $\Delta \mu_e(r, \, \theta) \equiv \mu_p+\mu_e-\mu_n$ 
and $\delta \mu_n^\infty(r,\,\theta)$ in the Legendre polynomials $P_l({\rm cos}\theta)$,
we demonstrate that all the components with $l \neq 0$
are \underline{fixed}
for stratified NSs
by specifying the magnetic field configuration.
This is in 
contrast to Refs.\ \cite{td96,hrv08,reisenegger09,hrv10,gjs11,bl16,papm17}, 
in which $\Delta \mu_e(r, \theta)$  
is determined from a single scalar differential equation 
depending on both the beta-reaction coefficient $\lambda_e$ and 
the relaxation 
timescale $\tau_{np}$.
	
\item The flow velocity ${\pmb U}$, defined by Eq.\ (\ref{222x}),
does not vanish and 
plays an important role in maintaining the quasi-equilibrium.
Its components $U_r$ and $U_\theta$ can be found 
from the continuity equations  (\ref{cont2p2}) and (\ref{cont2n2}), and
they depend, in particular, on the sources $\Delta \Gamma$.

\item The requirement of regularity of $U_r$ and $U_\theta$ at $r\rightarrow 0$,
$\theta \rightarrow 0$, and $\theta \rightarrow \pi$
allows us to determine the $l=0$
components of the functions 
$\Delta \mu_e$ and $\delta \mu_n^\infty$ (Appendix \ref{A3}).
It turns out that they depend on $T$
only in the narrow range of temperatures, where 
the dimensionless parameter 
$\lambda_e J_{np} R^2/n_p^2 \sim 1$.

\item The $\varphi$-component of the velocity ${\pmb U}$ 
is of special interest. 
It should be chosen in such a way to ensure
that the system is in the quasistationary state during its evolution 
[see the condition (\ref{uphi})].

\item The results listed above are obtained for $npe$ composition of NS cores
and for axisymmetric magnetic field configurations.
However, they can be easily generalized to include muons 
(and other particle species), 
non-axisymmetric magnetic fields, and superfluidity/superconductivity
(Secs.\ \ref{perturb5a}--\ref{perturb5c}).
They can also be generalized to the case when there are other factors
(in addition to the magnetic field)
disturbing the system from the diffusion and beta-equilibrium (Sec.\ \ref{deviations}).

\item We provide the formulas for the rate of magnetic field energy dissipation
for both normal $npe$ and $npe\mu$-matter 
[see Eqs.\ (\ref{EBdissnpe}) and (\ref{EBdissnpemu})].
These formulas retain its form in the superfluid matter, see Ref.\ \cite{kg17}.
In the limiting case when electron-neutron collisions are neglected ($J_{en}=0$),
our Eq.\ (\ref{EBdissnpe}) reduces to the well-known result of Ref.\ \cite{gr92}.
What is more interesting, we formulate a theorem 
which states that, under quasistationary conditions (see Sec.\ \ref{perturb2}), 
\underline{all} the heat generated in the system
is due to dissipation of the magnetic energy, excluding 
possible 
losses through the system boundary, 
$\dot{E}_{\rm B} = -\int_V T \dot{S} \,dV +{\rm ``Surface \, terms"}$.

\item Our results are illustrated by a numerical example
in which we calculate the dissipation timescales 
for the magnetic field
as functions of typical field 
and temperature (Sec.\ \ref{example}).
We demonstrate, in particular, that our ambipolar diffusion timescale
coincides with the solenoidal ambipolar timescale of Ref.\ \cite{gr92},
while the irrotational timescale (and the corresponding regime, 
see, e.g., Refs.\ \cite{gr92,td96,act04,hrv08,reisenegger09,dss09,hrv10,gjs11,dgp12,bl16,papm17}) 
does not appear in our analysis.

\item We see three immediate directions for future work.
First, it would be extremely interesting to calculate the flow velocity ${\pmb U}$
and hence to obtain all the necessary ingredients 
to follow the quasistationary magnetic field evolution in NSs.
Second, an important problem concerns the topology of currents in the vicinity 
of the crust-core interface. 
How much magnetic energy flows away from the core and dissipates in the crust?
This problem was completely ignored in the 
present paper.
Third, the present work indicates the need to re-examine magnetothermal evolution 
of NSs, especially, magnetars. 
Could the observed surface temperature of magnetars be supported 
by the magnetic field dissipation in their cores?
What is the role of suprathermal regime
($\Delta \mu_e \gtrsim k_{\rm B}T $) of beta-processes in such evolution?
We hope to address 
these issues
in our future work.

\end{itemize}

\begin{acknowledgments}
%
We are very grateful to A.I.~Chugunov, 
A.~Reisenegger, P.S.~Shternin, and D.G.~Yakovlev for useful discussions
and critical comments.
This work is supported in part by the Foundation for the advancement
of theoretical physics `BASIS' (grants No. 17-12-204-1 and 17-15-509-1). 
\end{acknowledgments}

\appendix

\section{Solution to Eq.\ (\ref{44444xx})}
\label{A1}

Introducing the parameters $Z_1$ and $Z_2$,
\begin{align}
&Z_1=n_e \, \Delta \mu_e + n_b \, \delta \mu_n^\infty,&
\label{Z1}\\
&Z_2 = -\left( \frac{dn_e}{dr} \,\Delta\mu_e
+\frac{dn_b}{dr} \, \delta \mu_n^\infty \right),&
\label{Z2}
\end{align}
Eq.\ (\ref{44444xx}) can be rewritten as %
%
\footnote{Similar equation has been recently discussed in Ref.\ \cite{gl16}.}
%
%
\begin{align}
{\pmb \nabla} Z_1 + Z_2 \, {\pmb e}_r = {\pmb F}_{\rm L},
\label{Eq2}
\end{align}
where ${\pmb F}_{\rm L}=[{\pmb j}\times {\pmb B}]/c$ 
is the Lorentz force density.
The solution to this equation reads
\begin{align}
&Z_1 = \int_0^\theta r F_{{\rm L}  \theta} \, d\tilde{\theta} + \zeta(r),&
\label{Z11}\\
&Z_2 =F_{{\rm L}  r}-\frac{\partial}{\partial r} 
\left[
\int_0^\theta r \, F_{{\rm L} \theta} \, d \tilde{\theta} 
\right]
- \zeta'(r),&
\label{Z22}
\end{align}
where the function $\zeta(r)$ is determined in Sec.\ \ref{perturb44a} 
(see also Appendix \ref{A3}).
Using Eqs.\ (\ref{Z11}) and (\ref{Z22}), one finds%
%
\footnote{This solution exists only for stratified stars. 
See also footnote \ref{nonstrat}.}
%
%
\begin{align}
&\left(
\begin{array}{c}
\Delta \mu_e \\
\delta \mu_n^\infty
\end{array}
\right)
=\left(
\begin{array}{cc}
\,\,\,\, n_e & \,\,\,\,n_b\\
-\frac{dn_e}{dr} & -\frac{dn_b}{dr}
\end{array}
\right)^{-1}
\left(
\begin{array}{c}
Z_1 \\
Z_2
\end{array}
\right).&
\label{solve}
\end{align}
Note that, as follows from 
this equation,
if we expand  $\Delta \mu_e$ (or $\delta \mu_n^\infty$) 
in the series of Legendre polynomials $P_l({\rm cos \theta})$, 
$\Delta \mu_e = \sum_l \Delta \mu_{e l} \, P_l({\rm cos}\theta)$,
then
the harmonics $\Delta \mu_{e l}$ 
with $l\neq 0$ will be independent 
of $\zeta(r)$, i.e., they are \underline{fully determined by the magnetic field configuration}.

Instead of the quantities $\Delta \mu_e$ and $\delta \mu_n^\infty$
it can be convenient to introduce the dimensionless parameters 
$\widetilde{\Delta \mu_e}$ and $\widetilde{\delta \mu_n^\infty}$
according to definitions
\begin{align}
&\Delta \mu_e \equiv \frac{\widetilde{B}^2}{\widetilde{n}}\, \widetilde{\Delta \mu_e},&
\label{dmue}\\
&\delta \mu_n^\infty \equiv \frac{\widetilde{B}^2}{\widetilde{n}}\, \widetilde{\delta \mu_n^\infty},&
\label{dmun}
\end{align}
where $\widetilde{B}$ and $\widetilde{n}$
are some typical values of $B$ and number densities, respectively. 
The dimensionless parameter $\widetilde{\delta \mu_n^\infty}$ 
is of the order of 
$\widetilde{n}/(4 \pi n_b)$
in the star, while
$\widetilde{\Delta \mu_e} \sim (n_b/n_e) \, \widetilde{\delta \mu_n^\infty}$
for the magnetic field configurations considered in this paper.

\section{Traditional derivation of the scalar equation for $\Delta \mu_e$}
\label{critique}

Here we present the ``traditional'' derivation of the scalar equation for $\Delta \mu_e$
following the recent work \cite{papm17}, 
and briefly discuss 
why (as we believe) the solution to this equation should not be relied upon.
Below we consider $npe$-matter (i.e., $n_e=n_p$)
and assume that $J_{en}=0$ (neutrons do not interact with electrons). 
Then, using Eqs.\ (\ref{4444xx}), (\ref{dm1}), and (\ref{neutron}),
one obtains
\begin{align}
{\pmb \nabla} \Delta \mu_e 
+\frac{\mu_p}{x_n^2 c^2 \tau_{pn}}\, {\pmb v}_{\rm amb}
=\frac{{\pmb F_{\rm L}}}{n_e},
\label{eq1}
\end{align}
where ${\pmb v}_{\rm amb} \equiv x_n ({\pmb u}_p-{\pmb u}_n)$; 
$x_n \equiv n_n/n_b$;
$\tau_{pn}=\mu_p n_p/(c^2 J_{np})$.
Taking divergence of this equation 
and using the continuity equations (\ref{cont2e})--(\ref{cont2n}),
one finds
\begin{align}
{\rm div} {\pmb \nabla} \Delta \mu_e -\frac{1}{b} \, \frac{\partial \Delta \mu_e}{\partial r}
-\beta \, {\rm div}(n_b \,{\pmb u}_n)={\rm div}
\left( \frac{{\pmb F}_{\rm L}}{n_e} \right)
-\frac{1}{b} \frac{F_{{\rm L}r}}{n_e},
\label{diveq}
\end{align}
where $1/b\equiv (1/\beta)\, d\beta/dr$ and $\beta \equiv \mu_p/(c^2 x_n n_p \tau_{pn})$.
The next step in the traditional approach consists in  
expressing ${\rm div}(n_b {\pmb u}_n)$ through $\Delta \mu_e$,
which requires some further approximations \cite{papm17}.
For example, one can write
(below we only consider the subthermal regime in which 
$\Delta \Gamma=\lambda_e \Delta \mu_e$;
see Secs.\ \ref{dissip} and \ref{example} for details)

\begin{align} 
x_n \, {\rm div} (n_b \, {\pmb u}_n) = x_n \, {\rm div} \left(\frac{{n_n} \, {\pmb u}_n}{x_n} \right)
= \lambda_e \Delta \mu_e -{\pmb \nabla}x_n \, (n_b\,  {\pmb u}_n)
\approx \lambda_e \Delta \mu_e,
\label{div}
\end{align}
where we, following Ref.\ \cite{papm17}, 
neglected the term proportional to ${\pmb \nabla}x_n$.
Using this approximation, Eq.\ (\ref{diveq}) takes the final form \cite{papm17}
\begin{align}
{\rm div} {\pmb \nabla} \Delta \mu_e -\frac{1}{b} \, \frac{\partial \Delta \mu_e}{\partial r}
-\frac{1}{a^2} \Delta \mu_e
={\rm div}
\left( \frac{{\pmb F}_{\rm L}}{n_e} \right)
-\frac{1}{b} \frac{F_{{\rm L}r}}{n_e},
\label{diveq1}
\end{align}
where $1/a^2\equiv \beta \lambda_e/x_n$.
The authors of Ref.\ \cite{papm17} impose the following boundary conditions 
for this equation:
regularity of the ambipolar velocity at the origin and the magnetic axis,
and vanishing of its radial component at the crust-core interface.%
%
\footnote{
Actually, we see no physical reason to require that the radial component of the 
ambipolar velocity vanishes at the crust-core interface:
nothing can prevent neutrons and protons from penetrating into the crust, 
where they can
suffer direct and inverse beta-decays, 
interact with the existing nuclei or form the new ones.
Of course, in the crust the dynamical equations 
for nucleons will differ from Eq.~(\ref{main1}).
}
%
With these boundary conditions Eq.\ (\ref{diveq1}) can be solved 
and it is easy to see that, generally, $\Delta \mu_e$
will depend on
the relaxation time $\tau_{pn}$ 
and the beta-reaction rate (through the coefficient $\lambda_e$).%
%
\footnote{It is important to stress that
\underline{all} harmonics in the expansion of $\Delta \mu_e$ 
in Legendre polynomials $P_l({\rm cos}\theta)$ 
will generally depend on $\tau_{pn}$ and $\lambda_e$.
This is in contrast to our solution (\ref{solve}), 
in which only the $l=0$ harmonic may depend on $\tau_{pn}$ and $\lambda_e$
through the function $\zeta(r)$.}
%
This result apparently contradicts our solution (see Appendix \ref{A1}).
Moreover, it follows from Eq.\ (\ref{diveq1}) that at large temperatures 
(when $1/a^2\rightarrow \infty$)
$\Delta \mu_e=0$,
while our solution (\ref{solve}) predicts that $\Delta \mu_e$ 
does not necessary vanish and is determined by the current magnetic field configuration
(which, of course, will evolve in time to smooth out deviations from chemical equilibrium 
-- but we do not consider 
the magnetic field dynamics in the present paper).

So, what is wrong with Eq.\ (\ref{diveq1}) and/or its solution?
First of all, an approximation of Eq.\ (\ref{div}), 
when one neglects 
the term $-{\pmb \nabla}x_n \, (n_b\,  {\pmb u}_n)$ in comparison to $\lambda_e \Delta\mu_e$
is unjustified, because ${\pmb u}_n$ diverges at $\nabla x_n \rightarrow 0$ 
[see Eqs.\ (\ref{divun}) and (\ref{unr}) and the footnote \ref{diverge}].
Second,
even if we take Eq.\ (\ref{diveq1}) for granted,
it is not proven that 
the solution to this \underline{scalar} equation
is, at the same time, the solution to the initial \underline{vector} equation (\ref{eq1}) 
[or Eq.\ (\ref{4444xx})];
our analysis shows that it is not the case.

\section{Magnetic field structure}
\label{A2}

We use the same axisymmetric model of the 
magnetic field as in Ref.~\cite{papm17} 
(see this reference for a detailed description and justification of the model). 
In spherical coordinates the magnetic field is given by
\begin{equation}
	\label{field:main}
	{\pmb B} = \frac{1}{r\sin\theta} \left( {\pmb\nabla}\mathcal{P}\times {\pmb e}_\varphi + \mathcal{T}{\pmb e}_\varphi \right),
\end{equation}
where ${\pmb e}_\varphi$ is the unit vector in the azimuthal direction; 
$\mathcal{P}(r,\theta)$ and $\mathcal{T}(r,\theta)$ 
are the poloidal and toroidal stream functions, respectively. 
They are expressed as
\begin{equation}
	\label{field:PandT}
	\mathcal{P} = \mathcal{P}_0 f(r/R_{\rm NS})\sin^2\theta, 
	\qquad 
	\mathcal{T} = \frac{s}{P_0 R_{\rm NS}} 
	\left( \mathcal{P}-\mathcal{P}_0 \right)^2 \Theta(\mathcal{P}-\mathcal{P}_0),
\end{equation}
where $R_{\rm NS}$ 
is the stellar radius;
$\Theta(x)$ is the Heaviside step function and the function 
\begin{equation}
	\label{field:f}
	f(x) = \begin{cases}
		\frac{35}{8}x^2 - \frac{21}{4}x^4 + \frac{15}{8}x^6, & x<1, \\
		\frac{1}{x},                                         & x\geqslant 1
	\end{cases}
\end{equation}
determines 
one of the possible 
polynomial configurations 
of the poloidal component, 
which is dipolar outside the star. 
One can check that this magnetic field configuration satisfies the condition (\ref{Lorentz}). 

This model is defined by two parameters, 
$\mathcal{P}_0$ and $s$. 
It is more convenient, however, to choose 
the maximum absolute value of the poloidal $B_{\rm P max}$ 
and toroidal $B_{\rm T max}$ components
as independent parameters.
They are related to $\mathcal{P}_0$ and $s$ by the formulas
\begin{equation}
	\label{field:Ps-BB}
	B_{\rm P max} = 8.75\frac{\mathcal{P}_0}{R_{\rm NS}^2}, \quad B_{\rm T max} \approx 0.0254\frac{s\mathcal{P}_0}{R_{\rm NS}^2}, \quad s \approx 345 \frac{B_{\rm T max}}{B_{\rm P max}}.
\end{equation}
In addition, there is a magnetic field $B_{\rm p}$ at the pole on the stellar surface. 
From Eqs.~(\ref{field:main})--(\ref{field:f}) it follows that 
$B_{\rm p} \approx 0.229 B_{\rm P max}$.
Note that, at fixed ratio $B_{\rm T max}/B_{\rm P max}$, 
the magnetic field configuration is determined by the only one scaling parameter, 
e.g., $B_{\rm P max}$ or $B_{\rm p}$.

\section{Calculation of $\zeta(r)$}
\label{A3}

The general scheme of Sec.\ \ref{perturb44a}
can be, of course, applied to the simple case of Sec.\ \ref{example}.
However, it is easier to slightly modify it in this particular situation.
Namely, it is convenient to start directly from the continuity equations (\ref{cont})
for protons and neutrons, which read, in the quasistationary approximation,
\begin{align}
&\frac{dn_p}{dr} u_{pr} + n_p \, {\rm div} \, {\pmb u}_p=-\Delta \Gamma,&
\label{p}\\
&\frac{dn_n}{dr} u_{nr} + n_n \, {\rm div} \, {\pmb u}_n= \Delta \Gamma,&
\label{n}
\end{align}
where the number densities $n_p$, $n_n$ depend on $r$ only and
$u_{pr}$, $u_{nr}$ are the radial components 
of the proton and neutron velocities ${\pmb u}_p$ and ${\pmb u}_n$, 
respectively.
Note that these velocities are not independent.
Assuming $J_{en}\ll J_{np}$, it follows from Eq.\ (\ref{neutron}) 
\begin{align}
&{\pmb u}_p={\pmb u}_n+ \frac{n_n}{J_{np}} \, {\pmb \nabla}\delta\mu_n^\infty.&
\label{up}
\end{align}
Using these equations, one can find 
${\rm div} \, {\pmb u}_n$ and $u_{nr}$%
%
\footnote{
\label{diverge}	
Note that these quantities diverge 
in non-stratified neutron stars, since $1/(y_n-y_p) \propto 1/(\nabla x_n)$, 
where $x_n=n_n/n_b$.}:
%
%
\begin{align}
&{\rm div}\, {\pmb u}_n = -\frac{1}{y_n-y_p}
\left[
y_n \, {\rm div} \, {\pmb M}_n+ y_n y_p \, M_{nr}
+ \left(\frac{y_p}{n_n} +\frac{y_n}{n_p} \right) \, \Delta \Gamma
\right],&
\label{divun}\\
&u_{nr} = \frac{1}{y_n-y_p} 
\left[ 
{\rm div} \,  {\pmb M}_n + y_p \, M_{nr} +\left( \frac{1}{n_n}+\frac{1}{n_p} \right) \Delta \Gamma
\right],&
\label{unr}
\end{align}
where 
\begin{align}
&y_i \equiv \frac{1}{n_i} \frac{dn_i}{dr}, \quad i=n, \, p,&
\label{xi}\\
&{\pmb M}_n \equiv \frac{n_n}{J_{np}} \, {\pmb \nabla}\delta \mu_n^\infty.&
\label{Mn}
\end{align}
Since $y_i \rightarrow O(r)$, at $r \rightarrow 0$, 
while $n_n$, $n_p$ and $J_{np} \rightarrow {\rm const}$ at $r \rightarrow 0$, 
it follows from Eq.\ (\ref{unr}) that $u_{nr}$ is finite at $r \rightarrow 0$
only if
\begin{align}
&\left[\frac{n_n}{J_{np}} \, \nabla^2(\delta \mu_n^\infty) +	
\left(\frac{1}{n_n}+\frac{1}{n_p} \right) \Delta \Gamma \right]_{r \rightarrow 0} 
\sim r^\alpha, \quad \alpha \geq 1.&
\label{r0dG}
\end{align}
Using Eqs.\ (\ref{divun}) and (\ref{unr}) we can find $u_{n\theta}$
and hence determine an analogue of the condition (\ref{Uthetacond}) 
ensuring finiteness of $u_{n\theta}$,%
%
\footnote{It is clear that if ${\pmb u}_n$ is well-behaved, then all other 
velocities (including ${\pmb U}$) are also well-behaved 
and can be easily expressed through~${\pmb u}_n$.}
%
%
\begin{align}
&\int_0^\pi r \, {\rm sin} \tilde{\theta} \left[ {\rm div} \, {\pmb u}_n - \frac{1}{r^2} 
\frac{\partial(r^2 u_{nr})}{\partial r}
\right] d \tilde{\theta}=0.&
\label{uthcond}
\end{align}
This condition can be conveniently rewritten in an operator form as
\begin{align}
&\hat{\bf{P}}_0 \left[ {\rm div} \, {\pmb u}_n - \frac{1}{r^2} 
\frac{\partial(r^2 u_{nr})}{\partial r}
\right]=0,&
\label{uthcond2}
\end{align}
where the operator $\hat{\bf{P}}_0$ extracts $l=0$ component in the Legendre expansion 
of an arbitrary function $f(r,\, \theta)=\sum_{l=0}^\infty f_l(r) P_l({\rm cos}\, \theta)$:
$\hat{\bf{P}}_0 f(r,\, \theta) \equiv 1/2 \, \int_0^\pi {\rm sin}\theta \,f(r,\, \theta) \, d\theta = f_0(r)$. 
Now, if we substitute Eqs.\ (\ref{divun})--(\ref{Mn}) into Eq.\ (\ref{uthcond2}),
we obtain a third-order linear differential equation depending on
$(\hat{\bf{P}}_0 \delta\mu_n^\infty)'''$, 
$(\hat{\bf{P}}_0 \delta\mu_n^\infty)''$, 
$(\hat{\bf{P}}_0 \delta\mu_n^\infty)'$, 
$(\hat{\bf{P}}_0 \Delta \Gamma)'$, and 
$\hat{\bf{P}}_0 \Delta \Gamma$, 
where the prime ($'$) means $d/dr$.
Schematically, it can be presented as
\begin{align}
&A_1(r) \,(\hat{\bf{P}}_0 \delta\mu_n^\infty)'''+
A_2(r) \,(\hat{\bf{P}}_0 \delta\mu_n^\infty)''+
A_3(r) \,(\hat{\bf{P}}_0 \delta\mu_n^\infty)'+
A_4(r) \, (\hat{\bf{P}}_0 \Delta \Gamma)'+
A_5(r) \, \hat{\bf{P}}_0 \Delta \Gamma=0,&
\label{diffeq}
\end{align}
where $A_1(r)$, $A_2(r)$, $A_3(r)$, $A_4(r)$, and $A_5(r)$ 
are some coefficients that can be easily determined from Eqs.\ (\ref{divun}) and (\ref{unr}).
Eq.\ (\ref{diffeq}) should be supplemented with the boundary conditions.
They can be found, in particular, 
from 
Eq.\ 
(\ref{r0dG}),
whose $l=0$ component is given by
\begin{align}
&\left[\frac{n_n}{J_{np}} \, \nabla^2( \hat{\bf{P}}_0\delta \mu_n^\infty) +	
\left(\frac{1}{n_n}+\frac{1}{n_p} \right) \hat{\bf{P}}_0 \Delta \Gamma \right]_{r \rightarrow 0} 
\sim r^\alpha, \quad \alpha \geq 1.&
\label{rodG1}
\end{align}
Note that, Eqs.\ (\ref{diffeq}) and (\ref{rodG1}) 
depend on two functions, $\hat{\bf{P}}_0\delta \mu_n^\infty$
and $\hat{\bf{P}}_0 \Delta \Gamma$. 
In fact, they are not independent and are related by the 
($l=0$) radial component of Eq.\ (\ref{4444xx}), 
in which $\delta \Delta \mu_e^\infty=\Delta \mu_e$ [see Eq.\ (\ref{dm1})],
\begin{align}
&(\hat{\bf{P}}_0 \delta \mu_n^\infty)' = \frac{1}{n_b} \, \hat{\bf{P}}_0 F_{{\rm L}r} 
-\frac{n_e}{n_b} \, (\hat{\bf{P}}_0 \Delta \mu_e)'. &
\label{P0dmun}
\end{align}
Since $\Delta\Gamma$ is known function of $\Delta \mu_e$ 
[see Eqs.\ (\ref{lambdaM}) and (\ref{lambdaD}) in Sec.\ \ref{example}],
and all the harmonics in the expansion of $\Delta \mu_e$ in Legendre polynomials
except for $l=0$ are specified by the magnetic field 
[see Eq.\ (\ref{solve})], 
the relation between $\hat{\bf{P}}_0 \Delta \mu_e$
and $\hat{\bf{P}}_0 \Delta \Gamma$ can be established
after rather tedious but straightforward calculations.
To simplify the subsequent presentation, 
below
we consider the subthermal regime,
$\Delta \mu_e \ll k_{\rm B} T$, 
in which $\Delta \Gamma = \lambda_e \Delta \mu_e$,
where $\lambda_e$ 
is the beta-reaction coefficient
that can be found from Eqs.\ (\ref{lambdaM}) or (\ref{lambdaD}).

\vspace{0.2 cm}
\noindent

{\bf Digression:} Before proceeding further, let us make 
a 
following 
comment. 
It is easy to demonstrate that 
the last two terms in Eq.\ (\ref{diffeq}) 
[and the last term in Eq.\ (\ref{rodG1})] 
can be neglected
(i.e., beta-processes are not important)
if $n_p^2/(R^2 J_{np}) \gg \lambda_e$ ($R$ is the typical lengthscale).
Then Eq.\ (\ref{diffeq}) becomes a homogeneous differential equation 
with the boundary condition 
$\nabla^2 (\hat{\bf{P}}_0\delta \mu_n^\infty)|_{r \rightarrow 0} \sim r^\alpha$ 
($\alpha\geq1$) [see Eq.\ (\ref{rodG1})], 
which results in 
$(\hat{\bf{P}}_0 \delta \mu_n^\infty)'|_{r\rightarrow 0}=0$ and
$(\hat{\bf{P}}_0 \delta \mu_n^\infty)''|_{r\rightarrow 0}=0$.
It has a unique solution, 
$(\hat{\bf{P}}_0 \delta \mu_n^\infty)'=0$, i.e.,
$\hat{\bf{P}}_0 \delta \mu_n^\infty(r)=C$.
The constant $C$ here is arbitrary; it specifies the central baryon number density 
of our perturbed NS model (following Ref.\ \cite{hartle67}, we prefer to define 
a stellar configuration by choosing 
central baryon density rather than the total number of baryons 
in the perturbed star). 
In what follows we assume $C=0$.
In the opposite limit, $n_p^2/(R^2 J_{np}) \ll \lambda_e$
(diffusion is not efficient),
similar consideration leads to the solution
$\hat{\bf{P}}_0 \Delta\Gamma(r)=0$,
which reduces to $\hat{\bf{P}}_0 \Delta \mu_e=0$ in the subthermal regime.
Irrespective of the limit,
knowledge of one function [$\hat{\bf{P}}_0 \delta \mu_n^\infty$ 
or $\hat{\bf{P}}_0 \Delta \mu_e$]
allows one to determine the derivative of another function 
using Eq.\ (\ref{P0dmun}).

As follows from these examples, the solution in both limits
is \underline{not sensitive} to the temperature or 
a particular dissipation mechanism.
However, an interplay of the dissipation mechanisms 
(ambipolar diffusion and non-equilibrium beta-processes)
determines a range of
transition temperatures, 
defined by the condition 
$n_p^2/(R^2 J_{np}) \sim \lambda_e$,
at which one asymptotic solution
transforms into another.
Since $\lambda_e$ is a strong function of temperature, 
the transition region is quite narrow.

\vspace{0.2 cm}
\noindent

Generally, to solve Eq.\ (\ref{diffeq}) with the boundary condition (\ref{rodG1})
one needs 
to express $(\hat{\bf{P}}_0 \delta \mu_n^\infty)'$ 
in these equations
through 
$\hat{\bf{P}}_0 F_{{\rm L}r}$ 
and  $(\hat{\bf{P}}_0 \Delta \mu_e)'$
using Eq.\ (\ref{P0dmun}).
The resulting inhomogeneous differential equation 
allows one to determine the function $\hat{\bf{P}}_0 \Delta \mu_e(r)$
(we remind that we assume
$\hat{\bf{P}}_0 \Delta \Gamma= \lambda_e\,  \hat{\bf{P}}_0 \Delta \mu_e$).
The boundary conditions for $\hat{\bf{P}}_0 \Delta \mu_e(r)$ 
depend on
the behaviour of the magnetic field at $r\rightarrow 0$
and follow from the analysis of (\ref{rodG1}):
\begin{align}
&(\hat{\bf{P}}_0\Delta \mu_e)'|_{r \rightarrow 0}=0,&
\label{cond1}\\
&(\hat{\bf{P}}_0\Delta \mu_e)''|_{r \rightarrow 0}=
\left(
\frac{1}{n_e}\, \hat{\bf{P}}_0 F_{{\rm L}r}
\right)'|_{r\rightarrow 0} +
\left(\frac{\lambda \, J_{np} \, n_b^2}{3 \, n_n^2 n_p^2}\,
\hat{\bf{P}}_0\Delta \mu_e \right)|_{r\rightarrow 0}
.&
\label{cond2}
\end{align}
We have here only two boundary conditions, while to solve
Eq.\ (\ref{diffeq}) we need, generally, one more condition 
which specifies $\hat{\bf{P}}_0 \Delta \mu_e$ at some $r$.%
%
\footnote{Note that this condition is not needed to find the solution of Eq.\ (\ref{diffeq})
in two limiting cases considered above, 
i.e., when $\lambda \rightarrow 0$ or $J_{np}\rightarrow \infty$.}
%
Presumably, this additional condition could be obtained by matching 
the solution of (\ref{diffeq}) with the solution of similar equation in the crust,
but we have not tried to perform such an analysis.
Let us only mention that
in an idealised (and unrealistic) situation in which $\lambda=0$ in the crust 
(i.e., when beta-processes are suppressed in the crust)
the condition for $\hat{\bf{P}}_0 \Delta \mu_e$
could follow from the requirement that there are no net flow of neutrons from the core to the crust
(otherwise, the quasistationarity condition would break down, since, by assumption, 
neutrons cannot be converted into protons in the crust).
This requirement means $\int_{\rm core} \lambda\,  (\hat{\bf{P}}_0\Delta \mu_e) \, r^2 dr=0$ 
[see Eq.\ (\ref{n})],
which gives us a third necessary condition to solve (\ref{diffeq}).

Assuming that the functions $\hat{\bf{P}}_0\Delta \mu_e$ and 
$\hat{\bf{P}}_0\delta \mu_n^\infty$ are already defined,
an unknown function $\zeta(r)$ can 
be found from the $l=0$ component
of equation (\ref{Z11}),
\begin{align}
&\zeta(r)=
\hat{\bf{P}}_0 Z_1 
- \hat{\bf{P}}_0 \int_0^\theta r\,  F_{{\rm L}\theta} \, d \tilde{\theta}.&
\label{zeta}
\end{align}
The functions $\delta \mu_n^\infty(r, \, \theta)$ and $\Delta \mu_e(r, \, \theta)$
can then be found from Eq.\ (\ref{solve}).
Alternatively, one can avoid use of the function $\zeta(r)$ 
by presenting the solution in the following equivalent way,
\begin{align}
&\delta \mu_n^\infty = (\delta \mu_n^\infty)_{\rm A6}
- \hat{\bf{P}}_0 (\delta \mu_n^\infty)_{\rm A6} 
+\hat{\bf{P}}_0 \delta \mu_n^\infty,&
\label{dmuninfty}\\
&\Delta \mu_e = (\Delta \mu_e)_{\rm A6}
- \hat{\bf{P}}_0 (\Delta \mu_e)_{\rm A6}
+\hat{\bf{P}}_0 \Delta \mu_e,&
\label{dmue2}
\end{align} 
where the functions $(\delta \mu_n^\infty)_{\rm A6}$ and $(\Delta \mu_e)_{\rm A6}$ 
are given by 
Eq.\ (\ref{solve}).
Although they depend on an (unknown) function $\zeta(r)$, 
one can choose this function in an arbitrary way  
(e.g., set $\zeta=0$) 
to calculate $(\delta\mu_{n}^{\infty})_{\rm A6}$ and $(\Delta\mu_{e})_{\rm A6}$, 
since it drops out from Eqs.\ (\ref{dmuninfty}) and (\ref{dmue2}). 

To plot Figs.\ \ref{fig:MU} and \ref{fig:DU}
we decided to use an approximate method for calculation of
$\zeta(r)$.
Namely, we employed the asymptotic solutions described above, 
assuming that 
$\hat{\bf{P}}_0 \delta \mu_n^\infty(r)=0$ 
in the ambipolar diffusion domain (see Figs.\ \ref{fig:MU} and \ref{fig:DU})
and 
$\hat{\bf{P}}_0 \Delta \mu_e=0$
in the MUrca (DUrca) domain.
We checked that the figures are not too
sensitive to an actual form of the solution.


\begin{thebibliography}{60}
	\expandafter\ifx\csname natexlab\endcsname\relax\def\natexlab#1{#1}\fi
	\expandafter\ifx\csname bibnamefont\endcsname\relax
	\def\bibnamefont#1{#1}\fi
	\expandafter\ifx\csname bibfnamefont\endcsname\relax
	\def\bibfnamefont#1{#1}\fi
	\expandafter\ifx\csname citenamefont\endcsname\relax
	\def\citenamefont#1{#1}\fi
	\expandafter\ifx\csname url\endcsname\relax
	\def\url#1{\texttt{#1}}\fi
	\expandafter\ifx\csname urlprefix\endcsname\relax\def\urlprefix{URL }\fi
	\providecommand{\bibinfo}[2]{#2}
	\providecommand{\eprint}[2][]{\url{#2}}
	
	\bibitem[{\citenamefont{{Kaspi}}(2010)}]{kaspi10}
	\bibinfo{author}{\bibfnamefont{V.~M.} \bibnamefont{{Kaspi}}},
	\bibinfo{journal}{Proceedings of the National Academy of Science}
	\textbf{\bibinfo{volume}{107}}, \bibinfo{pages}{7147} (\bibinfo{year}{2010}),
	\eprint{1005.0876}.
	
	\bibitem[{\citenamefont{{Vigan{\`o}} et~al.}(2013)\citenamefont{{Vigan{\`o}},
			{Rea}, {Pons}, {Perna}, {Aguilera}, and {Miralles}}}]{vigano_etal13}
	\bibinfo{author}{\bibfnamefont{D.}~\bibnamefont{{Vigan{\`o}}}},
	\bibinfo{author}{\bibfnamefont{N.}~\bibnamefont{{Rea}}},
	\bibinfo{author}{\bibfnamefont{J.~A.} \bibnamefont{{Pons}}},
	\bibinfo{author}{\bibfnamefont{R.}~\bibnamefont{{Perna}}},
	\bibinfo{author}{\bibfnamefont{D.~N.} \bibnamefont{{Aguilera}}},
	\bibnamefont{and} \bibinfo{author}{\bibfnamefont{J.~A.}
		\bibnamefont{{Miralles}}}, \bibinfo{journal}{\mnras}
	\textbf{\bibinfo{volume}{434}}, \bibinfo{pages}{123} (\bibinfo{year}{2013}),
	\eprint{1306.2156}.
	
	\bibitem[{\citenamefont{{Haensel} et~al.}(1990)\citenamefont{{Haensel},
			{Urpin}, and {Iakovlev}}}]{hui90}
	\bibinfo{author}{\bibfnamefont{P.}~\bibnamefont{{Haensel}}},
	\bibinfo{author}{\bibfnamefont{V.~A.} \bibnamefont{{Urpin}}},
	\bibnamefont{and} \bibinfo{author}{\bibfnamefont{D.~G.}
		\bibnamefont{{Iakovlev}}}, \bibinfo{journal}{\aap}
	\textbf{\bibinfo{volume}{229}}, \bibinfo{pages}{133} (\bibinfo{year}{1990}).
	
	\bibitem[{\citenamefont{{Urpin} and {Shalybkov}}(1995)}]{us95}
	\bibinfo{author}{\bibfnamefont{V.~A.} \bibnamefont{{Urpin}}} \bibnamefont{and}
	\bibinfo{author}{\bibfnamefont{D.~A.} \bibnamefont{{Shalybkov}}},
	\bibinfo{journal}{\aap} \textbf{\bibinfo{volume}{294}}, \bibinfo{pages}{117}
	(\bibinfo{year}{1995}).
	
	\bibitem[{\citenamefont{{Page} et~al.}(2000)\citenamefont{{Page}, {Geppert},
			and {Zannias}}}]{pgz00}
	\bibinfo{author}{\bibfnamefont{D.}~\bibnamefont{{Page}}},
	\bibinfo{author}{\bibfnamefont{U.}~\bibnamefont{{Geppert}}},
	\bibnamefont{and}
	\bibinfo{author}{\bibfnamefont{T.}~\bibnamefont{{Zannias}}},
	\bibinfo{journal}{\aap} \textbf{\bibinfo{volume}{360}}, \bibinfo{pages}{1052}
	(\bibinfo{year}{2000}), \eprint{astro-ph/0005301}.
	
	\bibitem[{\citenamefont{{Arras} et~al.}(2004)\citenamefont{{Arras}, {Cumming},
			and {Thompson}}}]{act04}
	\bibinfo{author}{\bibfnamefont{P.}~\bibnamefont{{Arras}}},
	\bibinfo{author}{\bibfnamefont{A.}~\bibnamefont{{Cumming}}},
	\bibnamefont{and}
	\bibinfo{author}{\bibfnamefont{C.}~\bibnamefont{{Thompson}}},
	\bibinfo{journal}{\apjl} \textbf{\bibinfo{volume}{608}}, \bibinfo{pages}{L49}
	(\bibinfo{year}{2004}), \eprint{astro-ph/0401561}.
	
	\bibitem[{\citenamefont{{Aguilera} et~al.}(2008)\citenamefont{{Aguilera},
			{Pons}, and {Miralles}}}]{apm08}
	\bibinfo{author}{\bibfnamefont{D.~N.} \bibnamefont{{Aguilera}}},
	\bibinfo{author}{\bibfnamefont{J.~A.} \bibnamefont{{Pons}}},
	\bibnamefont{and} \bibinfo{author}{\bibfnamefont{J.~A.}
		\bibnamefont{{Miralles}}}, \bibinfo{journal}{\aap}
	\textbf{\bibinfo{volume}{486}}, \bibinfo{pages}{255} (\bibinfo{year}{2008}),
	\eprint{0710.0854}.
	
	\bibitem[{\citenamefont{{Pons} et~al.}(2009)\citenamefont{{Pons}, {Miralles},
			and {Geppert}}}]{pmg09}
	\bibinfo{author}{\bibfnamefont{J.~A.} \bibnamefont{{Pons}}},
	\bibinfo{author}{\bibfnamefont{J.~A.} \bibnamefont{{Miralles}}},
	\bibnamefont{and}
	\bibinfo{author}{\bibfnamefont{U.}~\bibnamefont{{Geppert}}},
	\bibinfo{journal}{\aap} \textbf{\bibinfo{volume}{496}}, \bibinfo{pages}{207}
	(\bibinfo{year}{2009}), \eprint{0812.3018}.
	
	\bibitem[{\citenamefont{{Jones}}(1988)}]{jones88}
	\bibinfo{author}{\bibfnamefont{P.~B.} \bibnamefont{{Jones}}},
	\bibinfo{journal}{\mnras} \textbf{\bibinfo{volume}{233}},
	\bibinfo{pages}{875} (\bibinfo{year}{1988}).
	
	\bibitem[{\citenamefont{{Shalybkov} and {Urpin}}(1997)}]{su97}
	\bibinfo{author}{\bibfnamefont{D.~A.} \bibnamefont{{Shalybkov}}}
	\bibnamefont{and} \bibinfo{author}{\bibfnamefont{V.~A.}
		\bibnamefont{{Urpin}}}, \bibinfo{journal}{\aap}
	\textbf{\bibinfo{volume}{321}}, \bibinfo{pages}{685} (\bibinfo{year}{1997}).
	
	\bibitem[{\citenamefont{{Rheinhardt} and {Geppert}}(2002)}]{rg02}
	\bibinfo{author}{\bibfnamefont{M.}~\bibnamefont{{Rheinhardt}}}
	\bibnamefont{and}
	\bibinfo{author}{\bibfnamefont{U.}~\bibnamefont{{Geppert}}},
	\bibinfo{journal}{Physical Review Letters} \textbf{\bibinfo{volume}{88}},
	\bibinfo{eid}{101103} (\bibinfo{year}{2002}).
	
	\bibitem[{\citenamefont{{Hollerbach} and {R{\"u}diger}}(2004)}]{hr04}
	\bibinfo{author}{\bibfnamefont{R.}~\bibnamefont{{Hollerbach}}}
	\bibnamefont{and}
	\bibinfo{author}{\bibfnamefont{G.}~\bibnamefont{{R{\"u}diger}}},
	\bibinfo{journal}{\mnras} \textbf{\bibinfo{volume}{347}},
	\bibinfo{pages}{1273} (\bibinfo{year}{2004}).
	
	\bibitem[{\citenamefont{{Gourgouliatos}
			et~al.}(2013)\citenamefont{{Gourgouliatos}, {Cumming}, {Reisenegger},
			{Armaza}, {Lyutikov}, and {Valdivia}}}]{gcr_etal13}
	\bibinfo{author}{\bibfnamefont{K.~N.} \bibnamefont{{Gourgouliatos}}},
	\bibinfo{author}{\bibfnamefont{A.}~\bibnamefont{{Cumming}}},
	\bibinfo{author}{\bibfnamefont{A.}~\bibnamefont{{Reisenegger}}},
	\bibinfo{author}{\bibfnamefont{C.}~\bibnamefont{{Armaza}}},
	\bibinfo{author}{\bibfnamefont{M.}~\bibnamefont{{Lyutikov}}},
	\bibnamefont{and} \bibinfo{author}{\bibfnamefont{J.~A.}
		\bibnamefont{{Valdivia}}}, \bibinfo{journal}{\mnras}
	\textbf{\bibinfo{volume}{434}}, \bibinfo{pages}{2480} (\bibinfo{year}{2013}),
	\eprint{1305.6269}.
	
	\bibitem[{\citenamefont{{Gourgouliatos} and {Cumming}}(2014)}]{gc14}
	\bibinfo{author}{\bibfnamefont{K.~N.} \bibnamefont{{Gourgouliatos}}}
	\bibnamefont{and}
	\bibinfo{author}{\bibfnamefont{A.}~\bibnamefont{{Cumming}}},
	\bibinfo{journal}{Physical Review Letters} \textbf{\bibinfo{volume}{112}},
	\bibinfo{eid}{171101} (\bibinfo{year}{2014}), \eprint{1311.7345}.
	
	\bibitem[{\citenamefont{{Gourgouliatos}
			et~al.}(2016)\citenamefont{{Gourgouliatos}, {Wood}, and
			{Hollerbach}}}]{gwh16}
	\bibinfo{author}{\bibfnamefont{K.~N.} \bibnamefont{{Gourgouliatos}}},
	\bibinfo{author}{\bibfnamefont{T.~S.} \bibnamefont{{Wood}}},
	\bibnamefont{and}
	\bibinfo{author}{\bibfnamefont{R.}~\bibnamefont{{Hollerbach}}},
	\bibinfo{journal}{Proceedings of the National Academy of Science}
	\textbf{\bibinfo{volume}{113}}, \bibinfo{pages}{3944} (\bibinfo{year}{2016}),
	\eprint{1604.01399}.
	
	\bibitem[{\citenamefont{{Baym} et~al.}(1969)\citenamefont{{Baym}, {Pethick},
			and {Pines}}}]{bpp69}
	\bibinfo{author}{\bibfnamefont{G.}~\bibnamefont{{Baym}}},
	\bibinfo{author}{\bibfnamefont{C.}~\bibnamefont{{Pethick}}},
	\bibnamefont{and} \bibinfo{author}{\bibfnamefont{D.}~\bibnamefont{{Pines}}},
	\bibinfo{journal}{\nat} \textbf{\bibinfo{volume}{224}}, \bibinfo{pages}{674}
	(\bibinfo{year}{1969}).
	
	\bibitem[{\citenamefont{{Pethick}}(1992)}]{pethick92}
	\bibinfo{author}{\bibfnamefont{C.~J.} \bibnamefont{{Pethick}}}, in
	\emph{\bibinfo{booktitle}{Structure and Evolution of Neutron Stars}}, edited
	by \bibinfo{editor}{\bibfnamefont{D.}~\bibnamefont{{Pines}}},
	\bibinfo{editor}{\bibfnamefont{R.}~\bibnamefont{{Tamagaki}}},
	\bibnamefont{and} \bibinfo{editor}{\bibfnamefont{S.}~\bibnamefont{{Tsuruta}}}
	(\bibinfo{year}{1992}), p. \bibinfo{pages}{115}.
	
	\bibitem[{\citenamefont{{Urpin} and {Ray}}(1994)}]{ur94}
	\bibinfo{author}{\bibfnamefont{V.~A.} \bibnamefont{{Urpin}}} \bibnamefont{and}
	\bibinfo{author}{\bibfnamefont{A.}~\bibnamefont{{Ray}}},
	\bibinfo{journal}{\mnras} \textbf{\bibinfo{volume}{267}},
	\bibinfo{pages}{1000} (\bibinfo{year}{1994}).
	
	\bibitem[{\citenamefont{{Thompson} and {Duncan}}(1996)}]{td96}
	\bibinfo{author}{\bibfnamefont{C.}~\bibnamefont{{Thompson}}} \bibnamefont{and}
	\bibinfo{author}{\bibfnamefont{R.~C.} \bibnamefont{{Duncan}}},
	\bibinfo{journal}{\apj} \textbf{\bibinfo{volume}{473}}, \bibinfo{pages}{322}
	(\bibinfo{year}{1996}).
	
	\bibitem[{\citenamefont{{Urpin} and {Shalybkov}}(1999)}]{us99}
	\bibinfo{author}{\bibfnamefont{V.}~\bibnamefont{{Urpin}}} \bibnamefont{and}
	\bibinfo{author}{\bibfnamefont{D.}~\bibnamefont{{Shalybkov}}},
	\bibinfo{journal}{\mnras} \textbf{\bibinfo{volume}{304}},
	\bibinfo{pages}{451} (\bibinfo{year}{1999}).
	
	\bibitem[{\citenamefont{{Konenkov} and {Geppert}}(2000)}]{kg00}
	\bibinfo{author}{\bibfnamefont{D.}~\bibnamefont{{Konenkov}}} \bibnamefont{and}
	\bibinfo{author}{\bibfnamefont{U.}~\bibnamefont{{Geppert}}},
	\bibinfo{journal}{\mnras} \textbf{\bibinfo{volume}{313}}, \bibinfo{pages}{66}
	(\bibinfo{year}{2000}), \eprint{astro-ph/9910492}.
	
	\bibitem[{\citenamefont{{Konenkov} and {Geppert}}(2001)}]{kg01}
	\bibinfo{author}{\bibfnamefont{D.}~\bibnamefont{{Konenkov}}} \bibnamefont{and}
	\bibinfo{author}{\bibfnamefont{U.}~\bibnamefont{{Geppert}}},
	\bibinfo{journal}{\mnras} \textbf{\bibinfo{volume}{325}},
	\bibinfo{pages}{426} (\bibinfo{year}{2001}), \eprint{astro-ph/0103060}.
	
	\bibitem[{\citenamefont{{Braithwaite} and {Spruit}}(2006)}]{bs06}
	\bibinfo{author}{\bibfnamefont{J.}~\bibnamefont{{Braithwaite}}}
	\bibnamefont{and} \bibinfo{author}{\bibfnamefont{H.~C.}
		\bibnamefont{{Spruit}}}, \bibinfo{journal}{\aap}
	\textbf{\bibinfo{volume}{450}}, \bibinfo{pages}{1097} (\bibinfo{year}{2006}),
	\eprint{astro-ph/0510287}.
	
	\bibitem[{\citenamefont{{Hoyos} et~al.}(2008)\citenamefont{{Hoyos},
			{Reisenegger}, and {Valdivia}}}]{hrv08}
	\bibinfo{author}{\bibfnamefont{J.}~\bibnamefont{{Hoyos}}},
	\bibinfo{author}{\bibfnamefont{A.}~\bibnamefont{{Reisenegger}}},
	\bibnamefont{and} \bibinfo{author}{\bibfnamefont{J.~A.}
		\bibnamefont{{Valdivia}}}, \bibinfo{journal}{\aap}
	\textbf{\bibinfo{volume}{487}}, \bibinfo{pages}{789} (\bibinfo{year}{2008}),
	\eprint{0801.4372}.
	
	\bibitem[{\citenamefont{{Dall'Osso} et~al.}(2009)\citenamefont{{Dall'Osso},
			{Shore}, and {Stella}}}]{dss09}
	\bibinfo{author}{\bibfnamefont{S.}~\bibnamefont{{Dall'Osso}}},
	\bibinfo{author}{\bibfnamefont{S.~N.} \bibnamefont{{Shore}}},
	\bibnamefont{and} \bibinfo{author}{\bibfnamefont{L.}~\bibnamefont{{Stella}}},
	\bibinfo{journal}{\mnras} \textbf{\bibinfo{volume}{398}},
	\bibinfo{pages}{1869} (\bibinfo{year}{2009}), \eprint{0811.4311}.
	
	\bibitem[{\citenamefont{{Hoyos} et~al.}(2010)\citenamefont{{Hoyos},
			{Reisenegger}, and {Valdivia}}}]{hrv10}
	\bibinfo{author}{\bibfnamefont{J.~H.} \bibnamefont{{Hoyos}}},
	\bibinfo{author}{\bibfnamefont{A.}~\bibnamefont{{Reisenegger}}},
	\bibnamefont{and} \bibinfo{author}{\bibfnamefont{J.~A.}
		\bibnamefont{{Valdivia}}}, \bibinfo{journal}{\mnras}
	\textbf{\bibinfo{volume}{408}}, \bibinfo{pages}{1730} (\bibinfo{year}{2010}),
	\eprint{1003.5262}.
	
	\bibitem[{\citenamefont{{Ho}}(2011)}]{ho11}
	\bibinfo{author}{\bibfnamefont{W.~C.~G.} \bibnamefont{{Ho}}},
	\bibinfo{journal}{\mnras} \textbf{\bibinfo{volume}{414}},
	\bibinfo{pages}{2567} (\bibinfo{year}{2011}), \eprint{1102.4870}.
	
	\bibitem[{\citenamefont{{Dall'Osso} et~al.}(2012)\citenamefont{{Dall'Osso},
			{Granot}, and {Piran}}}]{dgp12}
	\bibinfo{author}{\bibfnamefont{S.}~\bibnamefont{{Dall'Osso}}},
	\bibinfo{author}{\bibfnamefont{J.}~\bibnamefont{{Granot}}}, \bibnamefont{and}
	\bibinfo{author}{\bibfnamefont{T.}~\bibnamefont{{Piran}}},
	\bibinfo{journal}{\mnras} \textbf{\bibinfo{volume}{422}},
	\bibinfo{pages}{2878} (\bibinfo{year}{2012}), \eprint{1110.2498}.
	
	\bibitem[{\citenamefont{{Graber} et~al.}(2015)\citenamefont{{Graber},
			{Andersson}, {Glampedakis}, and {Lander}}}]{gagl15}
	\bibinfo{author}{\bibfnamefont{V.}~\bibnamefont{{Graber}}},
	\bibinfo{author}{\bibfnamefont{N.}~\bibnamefont{{Andersson}}},
	\bibinfo{author}{\bibfnamefont{K.}~\bibnamefont{{Glampedakis}}},
	\bibnamefont{and} \bibinfo{author}{\bibfnamefont{S.~K.}
		\bibnamefont{{Lander}}}, \bibinfo{journal}{\mnras}
	\textbf{\bibinfo{volume}{453}}, \bibinfo{pages}{671} (\bibinfo{year}{2015}),
	\eprint{1505.00124}.
	
	\bibitem[{\citenamefont{{Elfritz} et~al.}(2016)\citenamefont{{Elfritz}, {Pons},
			{Rea}, {Glampedakis}, and {Vigan{\`o}}}}]{eprgv16}
	\bibinfo{author}{\bibfnamefont{J.~G.} \bibnamefont{{Elfritz}}},
	\bibinfo{author}{\bibfnamefont{J.~A.} \bibnamefont{{Pons}}},
	\bibinfo{author}{\bibfnamefont{N.}~\bibnamefont{{Rea}}},
	\bibinfo{author}{\bibfnamefont{K.}~\bibnamefont{{Glampedakis}}},
	\bibnamefont{and}
	\bibinfo{author}{\bibfnamefont{D.}~\bibnamefont{{Vigan{\`o}}}},
	\bibinfo{journal}{\mnras} \textbf{\bibinfo{volume}{456}},
	\bibinfo{pages}{4461} (\bibinfo{year}{2016}), \eprint{1512.07151}.
	
	\bibitem[{\citenamefont{{Beloborodov} and {Li}}(2016)}]{bl16}
	\bibinfo{author}{\bibfnamefont{A.~M.} \bibnamefont{{Beloborodov}}}
	\bibnamefont{and} \bibinfo{author}{\bibfnamefont{X.}~\bibnamefont{{Li}}},
	\bibinfo{journal}{\apj} \textbf{\bibinfo{volume}{833}}, \bibinfo{eid}{261}
	(\bibinfo{year}{2016}), \eprint{1605.09077}.
	
	\bibitem[{\citenamefont{{Shalybkov} and {Urpin}}(1995)}]{su95}
	\bibinfo{author}{\bibfnamefont{D.~A.} \bibnamefont{{Shalybkov}}}
	\bibnamefont{and} \bibinfo{author}{\bibfnamefont{V.~A.}
		\bibnamefont{{Urpin}}}, \bibinfo{journal}{\mnras}
	\textbf{\bibinfo{volume}{273}}, \bibinfo{pages}{643} (\bibinfo{year}{1995}).
	
	\bibitem[{\citenamefont{{Reisenegger}}(2009)}]{reisenegger09}
	\bibinfo{author}{\bibfnamefont{A.}~\bibnamefont{{Reisenegger}}},
	\bibinfo{journal}{\aap} \textbf{\bibinfo{volume}{499}}, \bibinfo{pages}{557}
	(\bibinfo{year}{2009}), \eprint{0809.0361}.
	
	\bibitem[{\citenamefont{{Glampedakis}
			et~al.}(2011{\natexlab{a}})\citenamefont{{Glampedakis}, {Jones}, and
			{Samuelsson}}}]{gjs11}
	\bibinfo{author}{\bibfnamefont{K.}~\bibnamefont{{Glampedakis}}},
	\bibinfo{author}{\bibfnamefont{D.~I.} \bibnamefont{{Jones}}},
	\bibnamefont{and}
	\bibinfo{author}{\bibfnamefont{L.}~\bibnamefont{{Samuelsson}}},
	\bibinfo{journal}{\mnras} \textbf{\bibinfo{volume}{413}},
	\bibinfo{pages}{2021} (\bibinfo{year}{2011}{\natexlab{a}}),
	\eprint{1010.1153}.
	
	\bibitem[{\citenamefont{{Passamonti} et~al.}(2017)\citenamefont{{Passamonti},
			{Akg{\"u}n}, {Pons}, and {Miralles}}}]{papm17}
	\bibinfo{author}{\bibfnamefont{A.}~\bibnamefont{{Passamonti}}},
	\bibinfo{author}{\bibfnamefont{T.}~\bibnamefont{{Akg{\"u}n}}},
	\bibinfo{author}{\bibfnamefont{J.~A.} \bibnamefont{{Pons}}},
	\bibnamefont{and} \bibinfo{author}{\bibfnamefont{J.~A.}
		\bibnamefont{{Miralles}}}, \bibinfo{journal}{\mnras}
	\textbf{\bibinfo{volume}{465}}, \bibinfo{pages}{3416} (\bibinfo{year}{2017}),
	\eprint{1608.00001}.
	
	\bibitem[{\citenamefont{{Goldreich} and {Reisenegger}}(1992)}]{gr92}
	\bibinfo{author}{\bibfnamefont{P.}~\bibnamefont{{Goldreich}}} \bibnamefont{and}
	\bibinfo{author}{\bibfnamefont{A.}~\bibnamefont{{Reisenegger}}},
	\bibinfo{journal}{\apj} \textbf{\bibinfo{volume}{395}}, \bibinfo{pages}{250}
	(\bibinfo{year}{1992}).
	
	\bibitem[{\citenamefont{{Iakovlev} and {Shalybkov}}(1991)}]{ys91a}
	\bibinfo{author}{\bibfnamefont{D.~G.} \bibnamefont{{Iakovlev}}}
	\bibnamefont{and} \bibinfo{author}{\bibfnamefont{D.~A.}
		\bibnamefont{{Shalybkov}}}, \bibinfo{journal}{Astrophys. Sp. Sci.}
	\textbf{\bibinfo{volume}{176}}, \bibinfo{pages}{171} (\bibinfo{year}{1991}).
	
	\bibitem[{\citenamefont{{Yakovlev} et~al.}(2001)\citenamefont{{Yakovlev},
			{Kaminker}, {Gnedin}, and {Haensel}}}]{ykgh01}
	\bibinfo{author}{\bibfnamefont{D.~G.} \bibnamefont{{Yakovlev}}},
	\bibinfo{author}{\bibfnamefont{A.~D.} \bibnamefont{{Kaminker}}},
	\bibinfo{author}{\bibfnamefont{O.~Y.} \bibnamefont{{Gnedin}}},
	\bibnamefont{and}
	\bibinfo{author}{\bibfnamefont{P.}~\bibnamefont{{Haensel}}},
	\bibinfo{journal}{\physrep} \textbf{\bibinfo{volume}{354}},
	\bibinfo{pages}{1} (\bibinfo{year}{2001}), \eprint{astro-ph/0012122}.
	
	\bibitem[{\citenamefont{{Landau} and {Lifshitz}}(1960)}]{ll60}
	\bibinfo{author}{\bibfnamefont{L.~D.} \bibnamefont{{Landau}}} \bibnamefont{and}
	\bibinfo{author}{\bibfnamefont{E.~M.} \bibnamefont{{Lifshitz}}},
	\emph{\bibinfo{title}{{Electrodynamics of continuous media}}}
	(\bibinfo{year}{1960}).
	
	\bibitem[{\citenamefont{{Braginskii}}(1965)}]{braginskii65}
	\bibinfo{author}{\bibfnamefont{S.~I.} \bibnamefont{{Braginskii}}}, in
	\emph{\bibinfo{booktitle}{Reviews of Plasma Physics (Consultants Bureau, New
			York NY)}}, edited by
	\bibinfo{editor}{\bibfnamefont{M.}~\bibnamefont{{Leontovich}}}
	(\bibinfo{year}{1965}), vol.~\bibinfo{volume}{1}, p. \bibinfo{pages}{205}.
	
	\bibitem[{\citenamefont{{Reisenegger}}(2007)}]{reisenegger07}
	\bibinfo{author}{\bibfnamefont{A.}~\bibnamefont{{Reisenegger}}},
	\bibinfo{journal}{Astronomische Nachrichten} \textbf{\bibinfo{volume}{328}},
	\bibinfo{pages}{1173} (\bibinfo{year}{2007}), \eprint{0710.2839}.
	
	\bibitem[{\citenamefont{{Glampedakis} and {Lasky}}(2016)}]{gl16}
	\bibinfo{author}{\bibfnamefont{K.}~\bibnamefont{{Glampedakis}}}
	\bibnamefont{and} \bibinfo{author}{\bibfnamefont{P.~D.}
		\bibnamefont{{Lasky}}}, \bibinfo{journal}{\mnras}
	\textbf{\bibinfo{volume}{463}}, \bibinfo{pages}{2542} (\bibinfo{year}{2016}),
	\eprint{1607.05576}.
	

	\bibitem[{\citenamefont{{Kantor} and {Gusakov}}(2017)}]{kg17}
	\bibinfo{author}{\bibfnamefont{E.~M.} \bibnamefont{{Kantor}}} \bibnamefont{and}
	\bibinfo{author}{\bibfnamefont{M.~E.} \bibnamefont{{Gusakov}}},
	\bibinfo{journal}{\mnras} \textbf{\bibinfo{volume}{473}},
	\bibinfo{pages}{4272} (\bibinfo{year}{2018}), 
	\eprint{1703.09216}.

	
	\bibitem[{\citenamefont{{Gusakov} and {Andersson}}(2006)}]{ga06}
	\bibinfo{author}{\bibfnamefont{M.~E.} \bibnamefont{{Gusakov}}}
	\bibnamefont{and}
	\bibinfo{author}{\bibfnamefont{N.}~\bibnamefont{{Andersson}}},
	\bibinfo{journal}{\mnras} \textbf{\bibinfo{volume}{372}},
	\bibinfo{pages}{1776} (\bibinfo{year}{2006}), \eprint{astro-ph/0602282}.
	
	\bibitem[{\citenamefont{{Gusakov}
			et~al.}(2009{\natexlab{a}})\citenamefont{{Gusakov}, {Kantor}, and
			{Haensel}}}]{gkh09a}
	\bibinfo{author}{\bibfnamefont{M.~E.} \bibnamefont{{Gusakov}}},
	\bibinfo{author}{\bibfnamefont{E.~M.} \bibnamefont{{Kantor}}},
	\bibnamefont{and}
	\bibinfo{author}{\bibfnamefont{P.}~\bibnamefont{{Haensel}}},
	\bibinfo{journal}{\prc} \textbf{\bibinfo{volume}{79}}, \bibinfo{eid}{055806}
	(\bibinfo{year}{2009}{\natexlab{a}}), \eprint{0904.3467}.
	
	\bibitem[{\citenamefont{{Gusakov}
			et~al.}(2009{\natexlab{b}})\citenamefont{{Gusakov}, {Kantor}, and
			{Haensel}}}]{gkh09b}
	\bibinfo{author}{\bibfnamefont{M.~E.} \bibnamefont{{Gusakov}}},
	\bibinfo{author}{\bibfnamefont{E.~M.} \bibnamefont{{Kantor}}},
	\bibnamefont{and}
	\bibinfo{author}{\bibfnamefont{P.}~\bibnamefont{{Haensel}}},
	\bibinfo{journal}{\prc} \textbf{\bibinfo{volume}{80}}, \bibinfo{eid}{015803}
	(\bibinfo{year}{2009}{\natexlab{b}}), \eprint{0907.0010}.
	
	\bibitem[{\citenamefont{{Gusakov}}(2016)}]{gusakov16}
	\bibinfo{author}{\bibfnamefont{M.~E.} \bibnamefont{{Gusakov}}},
	\bibinfo{journal}{\prd} \textbf{\bibinfo{volume}{93}}, \bibinfo{eid}{064033}
	(\bibinfo{year}{2016}), \eprint{1601.07732}.
	
	\bibitem[{\citenamefont{{Gusakov} and {Dommes}}(2016)}]{gd16}
	\bibinfo{author}{\bibfnamefont{M.~E.} \bibnamefont{{Gusakov}}}
	\bibnamefont{and} \bibinfo{author}{\bibfnamefont{V.~A.}
		\bibnamefont{{Dommes}}}, \bibinfo{journal}{\prd}
	\textbf{\bibinfo{volume}{94}}, \bibinfo{eid}{083006} (\bibinfo{year}{2016}),
	\eprint{1607.01629}.
	
	\bibitem[{\citenamefont{{Glampedakis}
			et~al.}(2011{\natexlab{b}})\citenamefont{{Glampedakis}, {Andersson}, and
			{Samuelsson}}}]{gas11}
	\bibinfo{author}{\bibfnamefont{K.}~\bibnamefont{{Glampedakis}}},
	\bibinfo{author}{\bibfnamefont{N.}~\bibnamefont{{Andersson}}},
	\bibnamefont{and}
	\bibinfo{author}{\bibfnamefont{L.}~\bibnamefont{{Samuelsson}}},
	\bibinfo{journal}{\mnras} \textbf{\bibinfo{volume}{410}},
	\bibinfo{pages}{805} (\bibinfo{year}{2011}{\natexlab{b}}),
	\eprint{1001.4046}.
	
	\bibitem[{\citenamefont{{Gusakov} and {Dommes}}(2017)}]{gd17}
	\bibinfo{author}{\bibfnamefont{M.~E.} \bibnamefont{{Gusakov}}}
	\bibnamefont{and} \bibinfo{author}{\bibfnamefont{V.~A.}
		\bibnamefont{{Dommes}}}, \bibinfo{journal}{in preparation}
	(\bibinfo{year}{2017}).
	
	\bibitem[{\citenamefont{{Yakovlev} and {Shalybkov}}(1990)}]{ys90}
	\bibinfo{author}{\bibfnamefont{D.~G.} \bibnamefont{{Yakovlev}}}
	\bibnamefont{and} \bibinfo{author}{\bibfnamefont{D.~A.}
		\bibnamefont{{Shalybkov}}}, \bibinfo{journal}{Soviet Astronomy Letters}
	\textbf{\bibinfo{volume}{16}}, \bibinfo{pages}{86} (\bibinfo{year}{1990}).
	
	\bibitem[{\citenamefont{{Yakovlev} and {Shalybkov}}(1991)}]{ys91b}
	\bibinfo{author}{\bibfnamefont{D.~G.} \bibnamefont{{Yakovlev}}}
	\bibnamefont{and} \bibinfo{author}{\bibfnamefont{D.~A.}
		\bibnamefont{{Shalybkov}}}, \bibinfo{journal}{\apss}
	\textbf{\bibinfo{volume}{176}}, \bibinfo{pages}{191} (\bibinfo{year}{1991}).
	
	\bibitem[{\citenamefont{{Reisenegger}}(1995)}]{reisenegger95}
	\bibinfo{author}{\bibfnamefont{A.}~\bibnamefont{{Reisenegger}}},
	\bibinfo{journal}{\apj} \textbf{\bibinfo{volume}{442}}, \bibinfo{pages}{749}
	(\bibinfo{year}{1995}), \eprint{astro-ph/9410035}.
	
	\bibitem[{\citenamefont{{Shternin}}(2008)}]{shternin08}
	\bibinfo{author}{\bibfnamefont{P.~S.} \bibnamefont{{Shternin}}},
	\bibinfo{journal}{Soviet Journal of Experimental and Theoretical Physics}
	\textbf{\bibinfo{volume}{107}}, \bibinfo{pages}{212} (\bibinfo{year}{2008}).
	
	\bibitem[{\citenamefont{{Heiselberg} and {Hjorth-Jensen}}(1999)}]{hhj99}
	\bibinfo{author}{\bibfnamefont{H.}~\bibnamefont{{Heiselberg}}}
	\bibnamefont{and}
	\bibinfo{author}{\bibfnamefont{M.}~\bibnamefont{{Hjorth-Jensen}}},
	\bibinfo{journal}{\apjl} \textbf{\bibinfo{volume}{525}}, \bibinfo{pages}{L45}
	(\bibinfo{year}{1999}), \eprint{astro-ph/9904214}.
	
	\bibitem[{\citenamefont{{Jones}}(2001)}]{jones01}
	\bibinfo{author}{\bibfnamefont{P.~B.} \bibnamefont{{Jones}}},
	\bibinfo{journal}{\prd} \textbf{\bibinfo{volume}{64}}, \bibinfo{eid}{084003}
	(\bibinfo{year}{2001}).
	
	\bibitem[{\citenamefont{{Lindblom} and {Owen}}(2002)}]{lo02}
	\bibinfo{author}{\bibfnamefont{L.}~\bibnamefont{{Lindblom}}} \bibnamefont{and}
	\bibinfo{author}{\bibfnamefont{B.~J.} \bibnamefont{{Owen}}},
	\bibinfo{journal}{\prd} \textbf{\bibinfo{volume}{65}}, \bibinfo{eid}{063006}
	(\bibinfo{year}{2002}), \eprint{astro-ph/0110558}.
	
	\bibitem[{\citenamefont{{Haensel} et~al.}(2002)\citenamefont{{Haensel},
			{Levenfish}, and {Yakovlev}}}]{hly02}
	\bibinfo{author}{\bibfnamefont{P.}~\bibnamefont{{Haensel}}},
	\bibinfo{author}{\bibfnamefont{K.~P.} \bibnamefont{{Levenfish}}},
	\bibnamefont{and} \bibinfo{author}{\bibfnamefont{D.~G.}
		\bibnamefont{{Yakovlev}}}, \bibinfo{journal}{\aap}
	\textbf{\bibinfo{volume}{381}}, \bibinfo{pages}{1080} (\bibinfo{year}{2002}),
	\eprint{astro-ph/0110575}.
	
	\bibitem[{\citenamefont{{Baiko} and {Yakovlev}}(1999)}]{by99}
	\bibinfo{author}{\bibfnamefont{D.~A.} \bibnamefont{{Baiko}}} \bibnamefont{and}
	\bibinfo{author}{\bibfnamefont{D.~G.} \bibnamefont{{Yakovlev}}},
	\bibinfo{journal}{\aap} \textbf{\bibinfo{volume}{342}}, \bibinfo{pages}{192}
	(\bibinfo{year}{1999}), \eprint{astro-ph/9812071}.
	
	\bibitem[{\citenamefont{{Hartle}}(1967)}]{hartle67}
	\bibinfo{author}{\bibfnamefont{J.~B.} \bibnamefont{{Hartle}}},
	\bibinfo{journal}{\apj} \textbf{\bibinfo{volume}{150}}, \bibinfo{pages}{1005}
	(\bibinfo{year}{1967}).
	
\end{thebibliography}

\end{document}